\documentclass[12pt,reqno]{amsart}

\usepackage{epsfig}
\usepackage{amsmath, amsthm, amsfonts}
\usepackage{graphicx}

\numberwithin{equation}{section}

\newcommand{\R}{{\mathbb R}}
\newcommand{\C}{{\mathbb C}}

\renewcommand{\Re}{{\operatorname{Re\,}}}
\renewcommand{\Im}{{\operatorname{Im\,}}}

\newcommand{\Tr}{{{\operatorname{Tr}}}}

\newcommand{\diag}{{\operatorname{diag}}}
\newcommand{\dist}{{\operatorname{dist}}}
\newcommand{\supp}{{\operatorname{supp}}}
\newcommand{\Ai}{{\operatorname{Ai}}}

\newcommand{\ga}{\gamma}
\newcommand{\Ga}{\Gamma}

\newcommand{\la}{\lambda}
\newcommand{\ep}{\varepsilon}

\newcommand{\f}{\varphi}

\newcommand{\ds}{\displaystyle}
\newtheorem{theo}{{\sc \bf Theorem}}[section]
\newtheorem{cor}[theo]{{\sc \bf Corollary}}
\newtheorem{lem}[theo]{{\sc \bf Lemma}}
\newtheorem{prop}[theo]{{\sc \bf Proposition}}

\newenvironment{defin}{\medskip\noindent{\it Definition:\/} }{\medskip}

\sc
\title[Large $n$ limit of Gaussian random matrices with external
source. Part III]
{Large $n$ limit of Gaussian random matrices with external
  source, Part III: Double scaling limit}
\author{Pavel M. Bleher}
\address{Department of Mathematical Sciences, Indiana
University-Purdue University Indianapolis,
402 N. Blackford St., Indianapolis, IN 46202, U.S.A.}
\email{bleher@math.iupui.edu}
\author{Arno B.J. Kuijlaars}
\address{Department of Mathematics,
Katholieke Universiteit Leuven, Celestijnenlaan 200 B, B-3001
Leuven, BELGIUM}
\email{arno.kuijlaars@wis.kuleuven.be}

\date{\today}

\thanks{The first author was supported in part by the National Science
  Foundation (NSF)
Grant DMS-0354962. The second author was supported by FWO-Flanders project
G.0455.04, by K.U. Leuven research grant OT/04/24,
by a grant from the Ministry of Education and Science
of Spain, project code MTM2005-08648-C02-01, and by
the European Science Foundation Program MISGAM}

\rm

\begin{document}

\begin{abstract}
We consider the double scaling limit in the random matrix ensemble with
an external source
\[ \frac{1}{Z_n} e^{-n \Tr(\frac{1}{2}M^2 -AM)} dM \]
defined on $n\times n$ Hermitian matrices, where $A$ is a diagonal
matrix with  two eigenvalues $\pm a$ of  equal multiplicities. The value
$a=1$ is critical since the eigenvalues of $M$ accumulate as $n \to \infty$
on two intervals for $a > 1$ and on one interval for $0 < a < 1$. These two
cases were treated in Parts I and II, where we showed that the local eigenvalue
correlations have the universal limiting behavior known from unitary random matrix
ensembles. For the critical case $a=1$ new limiting behavior occurs which is
described in terms of Pearcey integrals, as shown by Br\'ezin and Hikami,
and Tracy and Widom.  We establish this result by applying the Deift/Zhou
steepest descent method to a $3 \times 3$-matrix valued Riemann-Hilbert problem
which involves the construction of a local parametrix out of Pearcey integrals.
We resolve the main technical issue of matching the local Pearcey parametrix
with a global outside parametrix by modifying an underlying Riemann surface.
\end{abstract}

\maketitle

\section{Introduction and statement of results}

\subsection{The random matrix model}
\begin{figure}
\scalebox{0.4}{\includegraphics{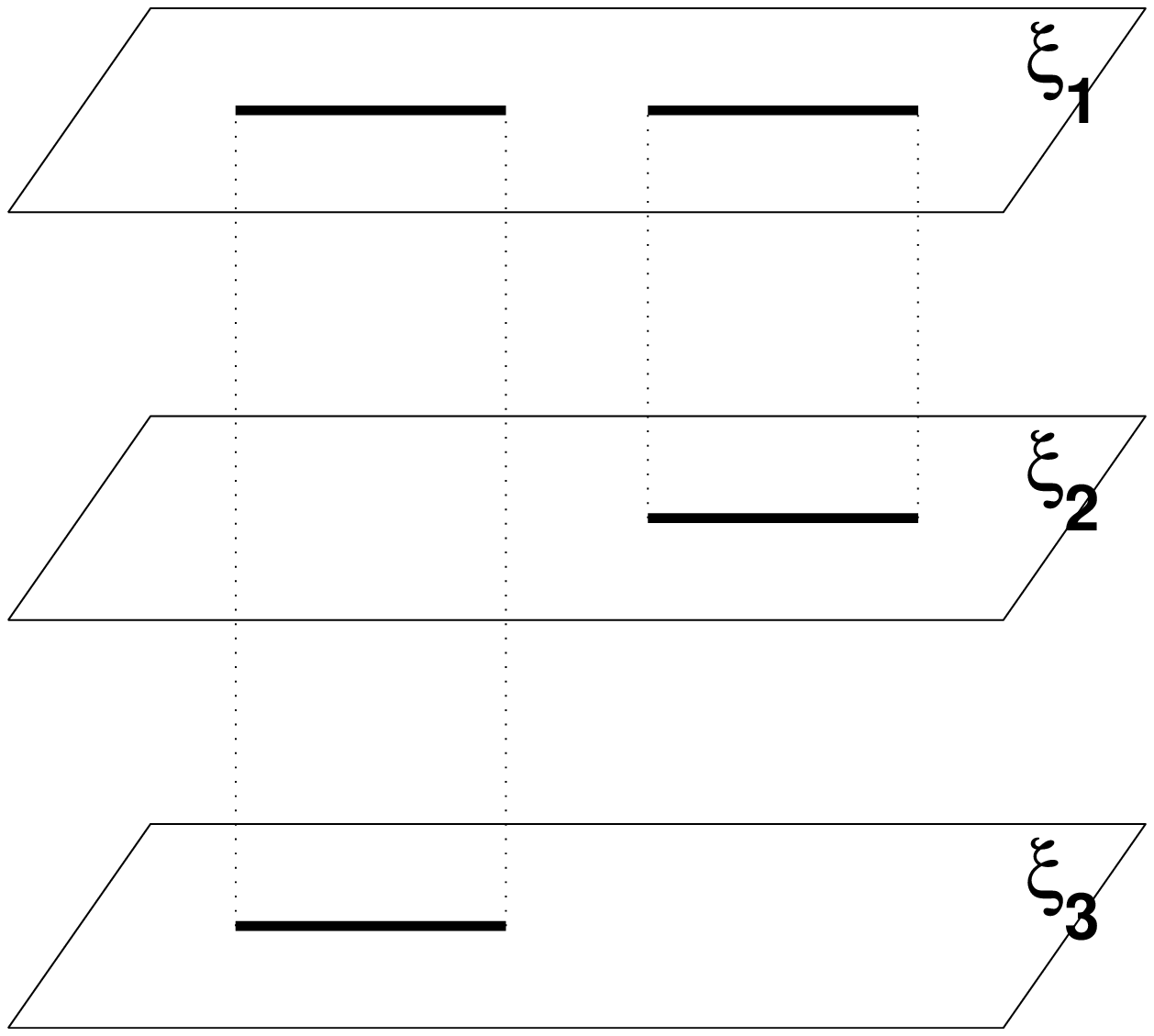}}
\scalebox{0.4}{\includegraphics{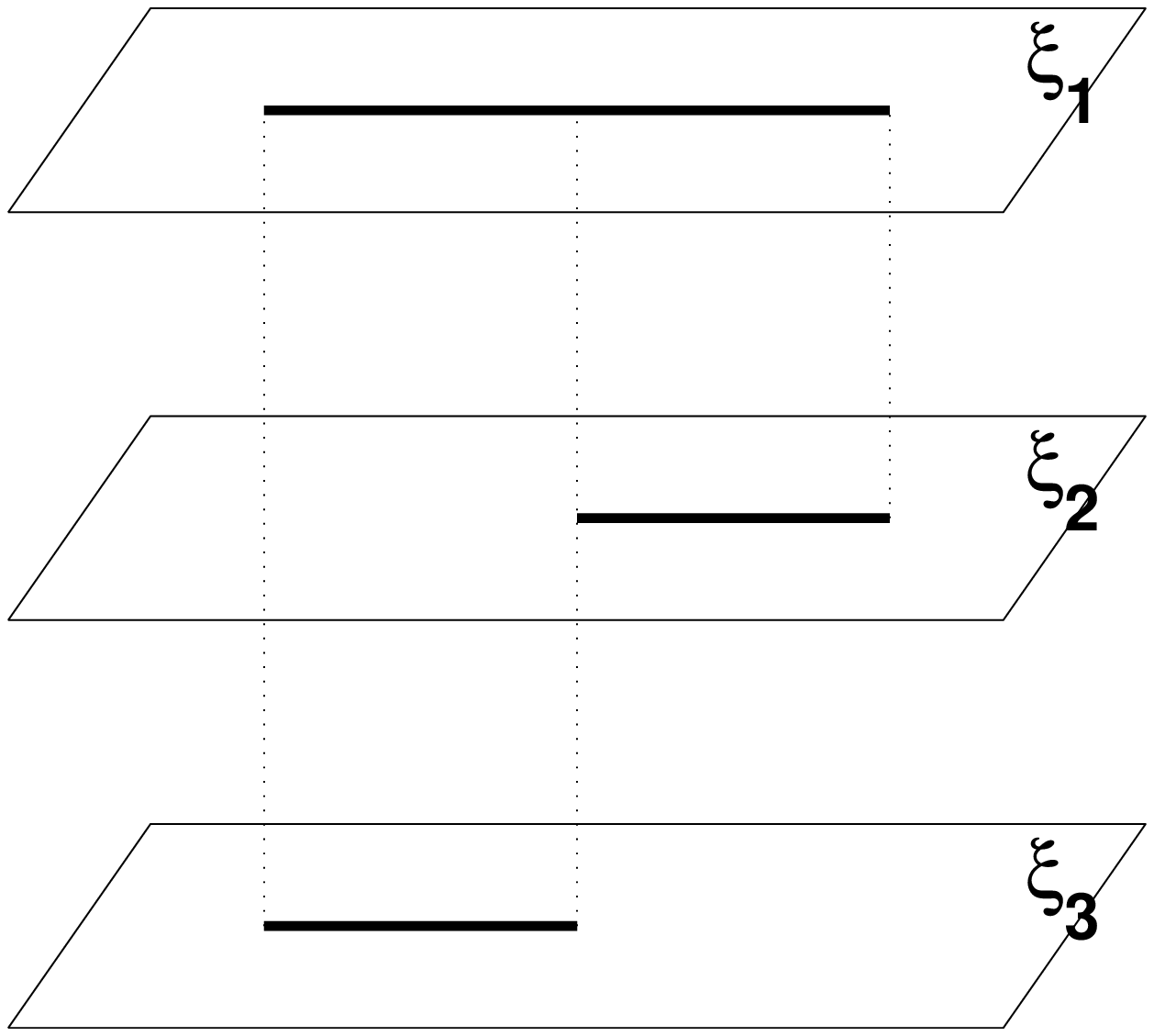}}
\scalebox{0.4}{\includegraphics{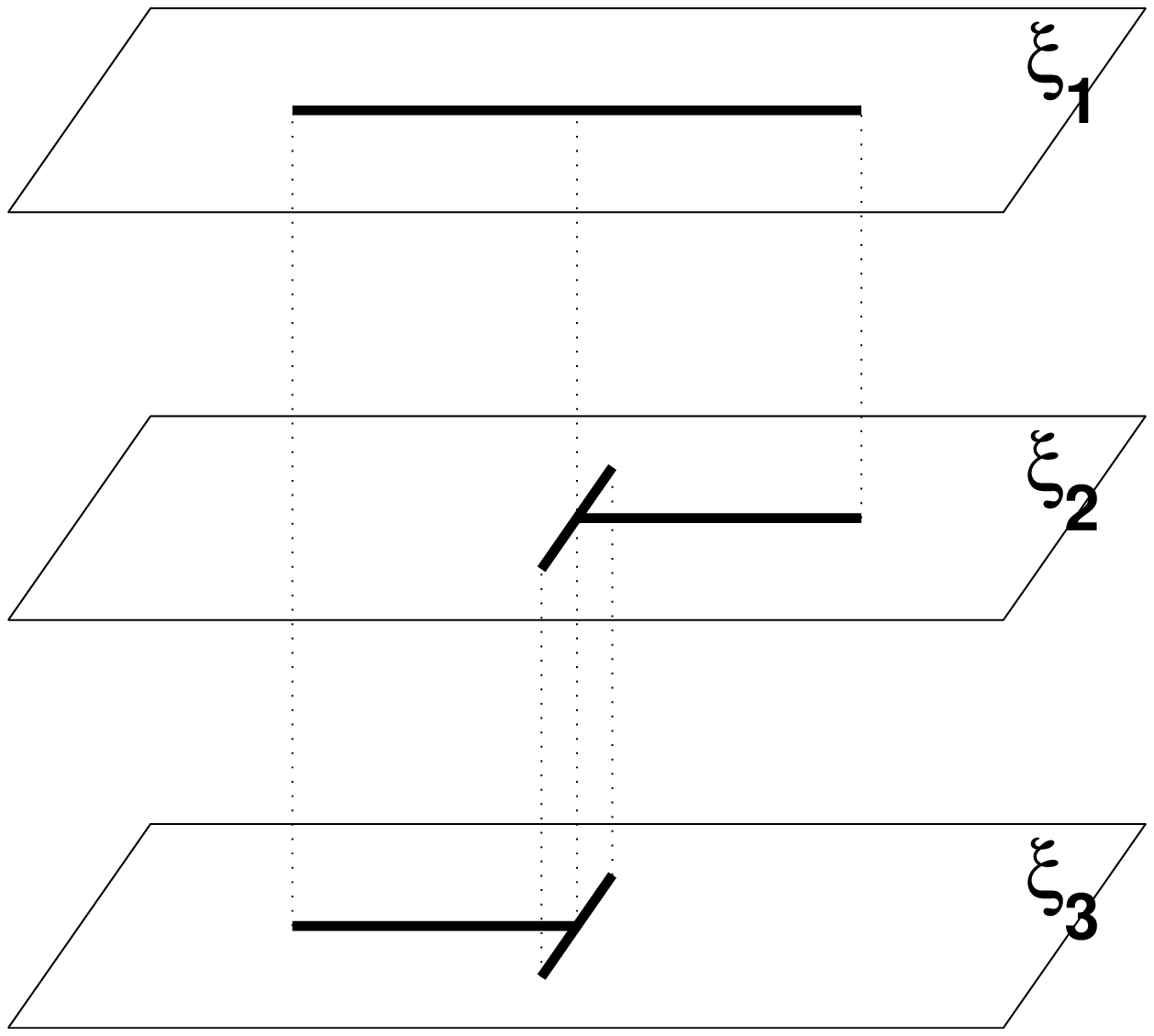}}
\caption{The structure of the Riemann surface for the equation (\ref{PasturEquation})
for the values $a> 1$ (left), $a=1$ (middle) and $a<1$ (right).
In all cases the eigenvalues of $M$ accumulate on the interval(s)
of the first sheet with a density given by (\ref{eigdensity}).}
\label{fig1}
\end{figure}

This is the third and final part of a sequence of papers on the Gaussian
random matrix ensemble with external source
\begin{equation} \label{matrixmodel}
 \frac{1}{Z_n} e^{-n \Tr(\frac{1}{2}M^2 -AM)} dM,
\end{equation}
defined on $n\times n$ Hermitian matrices, where $A$ is a diagonal
matrix with two eigenvalues $\pm a$ (with $a > 0$) of equal multiplicities
(so that, $n$ is even). This matrix ensemble was introduced
by Br\'ezin and Hikami \cite{BH1,BH2} as a simple model for a phase
transition that is expected to exhibit universality properties. The
phase transition can be seen from the behavior of the eigenvalues of $M$
in the large $n$ limit, since for $a > 1$, the eigenvalues accumulate
on two intervals, while for $0 < a < 1$, the eigenvalues accumulate on
one interval. The limiting mean density of eigenvalues  follows
from earlier work of Pastur \cite{Pas}.
It is based on an analysis of the equation (Pastur equation)
\begin{equation} \label{PasturEquation}
    \xi^3 - z \xi^2 + (1-a^2) \xi + a^2 z = 0,
\end{equation}
which yields an algebraic function $\xi(z)$ defined on a
three-sheeted Riemann surface. The restrictions of $\xi(z)$
to the three sheets are denoted by $\xi_j(z)$, $j=1,2,3$. There are four
real branch points if $a > 1$ which determine two real intervals.
The two intervals come together for $a=1$, and for $0 < a < 1$, there are
two real branch points, and two purely imaginary branch points.
Figure \ref{fig1} depicts the structure of the
Riemann surface $\xi(z)$ for $a>1$, $a=1$, and $a<1$.

In all cases we have that the limiting mean eigenvalue density
$\rho(x) = \rho(x;a)$ of the matrix $M$ from (\ref{matrixmodel})
is given by
\begin{equation} \label{eigdensity}
    \rho(x;a) = \frac{1}{\pi} \Im \xi_{1+}(x), \qquad x \in \mathbb R,
\end{equation}
where $\xi_{1+}(x)$ denotes the limiting value of $\xi_1(z)$ as $z \to x$
with $\Im z > 0$.
For $a = 1$ the limiting mean eigenvalue density vanishes
at $x=0$ and $\rho(x;a) \sim |x|^{1/3}$ as $x \to 0$.

We note that this behavior at the closing (or opening) of a gap
is markedly different from the behavior that occurs in the usual
unitary random matrix ensembles $Z_n^{-1} e^{-n \Tr V(M)} dM$ where a closing of
the gap in the spectrum typically leads to a limiting mean eigenvalue
density $\rho$ that satisfies $\rho(x) \sim (x-x^*)^2$ as $x \to x^*$
if the gap closes at $x= x^*$. In that case the local eigenvalue correlations
can be described in terms of $\psi$-functions associated with the
Painlev\'e II equation, see \cite{BI,CK}.
The phase transition for the model (\ref{matrixmodel}) is different,
and it cannot be realized in a unitary random matrix ensemble.

\subsection{Non-intersecting Brownian motion}
\begin{figure}[ht]
\scalebox{0.5}{\includegraphics{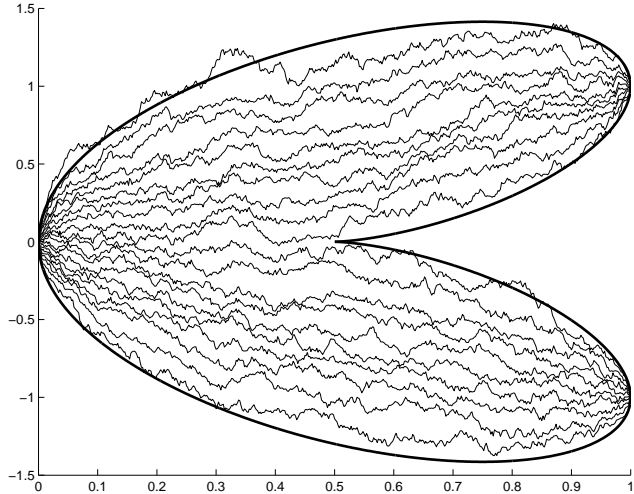}}
\caption{Non-intersecting Brownian paths that start at one point and end
at two points. At any intermediate time the positions of the paths are
distributed as the eigenvalues of a Gaussian random matrix ensemble with external
source. As their number increases the paths fill out a region whose boundary has a cusp.}
\label{figpaths}
\end{figure}

The nature of the phase transition at $a=1$ may also be seen from
an equivalent model of non-intersecting Brownian paths, see Figure \ref{figpaths}.
Consider $n$ independent
one-dimensional Brownian motions that are conditioned to start
at time $t=0$ at the origin, end at time $t=1$ at $\pm 1$,
where half of the paths ends at $+1$ and the other half at $-1$,
and that are conditioned not to intersect at intermediate times
$t \in (0,1)$.  As explained in \cite{ABK}, at any intermediate
time $t$, the positions of the $n$ Brownian motions, have the
same distribution as the eigenvalues of a Gaussian random matrix ensemble
with external source (up to trivial scaling).
Now, as $n \to \infty$ and under appropriate scaling
of the variance of the Brownian motions, the paths fill out
a region in the $t-x$-plane. Then for small time the paths are
in one group, which at a certain critical time $t_{cr}$
splits into two groups, where one group ends at $x=+1$ and the other
group at $x=-1$. The situations $t < t_{cr}$, $t = t_{cr}$, and $t > t_{cr}$
correspond to $a < 1$, $a=1$, and $a > 1$, respectively,
in the Gaussian random matrix model with external source.

The boundary curve has a cusp singularity at the critical time
as shown in Figure \ref{figpaths}.

\subsection{Correlation kernel}

Br\'ezin and Hikami \cite{BH1,BH2} showed, see also \cite{ZJ1},
that the eigenvalues of the random matrix ensemble
(\ref{matrixmodel})
are distributed according to a determinantal point process.
There is a kernel $K_n(x,y;a)$ so that the eigenvalues $x_1, \ldots, x_n$
have the joint probability density
\[ p_n(x_1, \ldots, x_n) = \frac{1}{n!} \det(K_n(x_j,x_k;a))_{j,k=1,\ldots, n} \]
and so that for each $m \leq n$, the  $m$-point correlation function
\[ R_m(x_1, \ldots, x_m) = \frac{n!}{(n-m)!}
    \underbrace{\int \cdots \int}_{n-m \textrm{ times}} p_n(x_1, \ldots, x_n) dx_{m+1} \cdots d x_n \]
takes determinantal form as well:
\[ R_m(x_1, \ldots, x_m) = \det (K_n(x_j,x_k;a))_{j,k=1,\ldots,m}. \]

In \cite{BK1} we pointed out that the kernel can be built out of
multiple Hermite polynomials \cite{ABVA,BK3,VAC} in much the same way that the correlation kernel
for unitary random matrix ensembles (without external source) is related to
orthogonal polynomials. The Christoffel-Darboux formula for multiple
orthogonal polynomials \cite{BK1,DK} allows one to express the kernel in terms
of the Riemann-Hilbert problem for multiple Hermite polynomials
(see \cite{VAGK} and below).
Applying the Deift/Zhou steepest descent analysis \cite{Dei,DZ}
to the Riemann-Hilbert problem in the non-critical case, we were able
to show that the kernel has
the usual scaling limits from random matrix theory. That is, we obtain the sine kernel
\begin{equation} \label{Kbulk}
    K^{bulk}(x,y) = \frac{\sin \pi(x-y)}{\pi(x-y)}
\end{equation}
 in the bulk, and the Airy kernel
\begin{equation} \label{Kedge}
    K^{edge}(x,y) = \frac{\Ai(x) \Ai'(y)-\Ai'(x) \Ai(y)}{x-y}
\end{equation}
at the edge of the spectrum,
as scaling limits of $K_n(x,y;a)$ if $a > 1$ \cite{BK2} or $0 < a < 1$ \cite{ABK}.

\subsection{Double scaling limit}
In this paper we  consider the double scaling limit at the critical
parameter $a=1$ of the Gaussian random matrix ensemble with external source,
or equivalently, of the non-intersecting Brownian motion model at
the critical time $t = t_{cr}$.
As is usual in a critical case, there is a family of limiting kernels
that arise when $a$ changes with $n$ and $a \to 1$ as $n \to \infty$
in a critical way. These kernels are constructed out of Pearcey integrals
and therefore they are called Pearcey kernels.
The Pearcey kernels were first described by Br\'ezin and Hikami \cite{BH1,BH2}.
A detailed proof of the following result was recently given by
Tracy and Widom \cite{TW}.

\begin{figure}[t]
\centerline{\includegraphics[height=5cm,width=5cm]{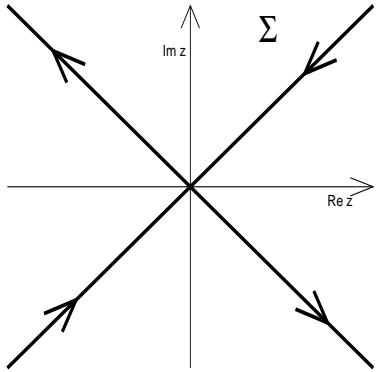}}
\caption{The contour $\Sigma$ that appears in the definition of $q(y)$.}
\label{fig:contourSigma}
\end{figure}

\begin{theo} \label{maintheo}
We have for every fixed $b \in \mathbb R$,
\begin{equation} \label{limitkernel}
    \lim_{n \to \infty} \frac{1}{n^{3/4}} K_n \left( \frac{x}{n^{3/4}}, \frac{y}{n^{3/4}};  1 + \frac{b}{2\sqrt{n}} \right)
    = K^{cusp}(x,y;b)
\end{equation}
where $K^{cusp}$ is the Pearcey kernel
\begin{equation} \label{Pearceykernel}
    K^{cusp}(x,y;b) =
    \frac{p(x) q''(y) - p'(x) q'(y) + p''(x) q(y) - bp(x) q(y)}{x-y}
       \end{equation}
with
\begin{equation} \label{Pearceyintegrals}
    p(x) = \frac{1}{2\pi} \int_{-\infty}^{\infty} e^{-\frac{1}{4}s^4 - \frac{b}{2} s^2 + isx} ds
    \qquad \mbox{ and } \qquad
   q(y) = \frac{1}{2\pi} \int_{\Sigma} e^{\frac{1}{4} t^4 + \frac{b}{2} t^2 + ity} dt.
   \end{equation}
The contour $\Sigma$ consists of the four rays $\arg y = \pm \pi/4, \pm 3\pi/4$,
with the orientation shown in Figure {\rm\ref{fig:contourSigma}}.
\end{theo}

The functions (\ref{Pearceyintegrals}) are called Pearcey integrals \cite{Pearcey}.
They are solutions of the third order differential equations $p'''(x) = xp(x) + bp'(x)$
and $q'''(y) = - yq(y) + bq'(y)$, respectively.
Away from the critical point $x=0$, the usual scaling limits (\ref{Kbulk}) and (\ref{Kedge})
from random matrix theory continue to hold in the case $a=1$ (also in
the double scaling regime).
This can be proved for example as in \cite{BK2,TW}, and we will not
consider this any further here.

Theorem \ref{maintheo} implies that local eigenvalue statistics of eigenvalues near $0$
are expressed in terms of the Pearcey kernel. For example we have the following
corollary of Theorem \ref{maintheo}.
\begin{cor}
The probability that a matrix of the ensemble {\rm (\ref{matrixmodel})}
with $a = 1 + bn^{-1/2}/2$ has no eigenvalues in the interval
$[c n^{-3/4}, d n^{-3/4}]$ converges, as $n \to \infty$,
to the Fredholm determinant of the integral operator with kernel
$K^{cusp}(x,y;b)$ acting on $L^2(c,d)$.
\end{cor}
Similar expressions hold for the probability to have one, two, three, \ldots, eigenvalues
in an $O(n^{-3/4})$ neighborhood of $x=0$.

Tracy and Widom \cite{TW} and Adler and van Moerbeke \cite{AvM} gave
differential equations for the gap probabilities associated
with the Pearcey kernel and with the more general Pearcey process
which arises from considering the non-intersecting Brownian motion
model at several times near the critical time. See also \cite{OR}
where the Pearcey process appears in a combinatorial model on
random partitions.

\subsection{Steepest descent method for RH problems}

Br\'ezin and Hikami and also Tracy and Widom used a double integral representation
for the kernel in order to establish Theorem \ref{maintheo}. In this paper we use
the Deift/Zhou steepest descent method for the Riemann-Hilbert problem for multiple Hermite polynomials.
This method is less direct than the steepest descent method for integrals. However,
an approach based on the Riemann-Hilbert problem may be applicable to more general situations,
where an integral representation is not available. This is the case, for example,
for the general (non-Gaussian) unitary random matrix ensemble with external source
\begin{equation} \label{generalmatrixmodel}
    \frac{1}{Z_n} e^{-n \Tr (V(M) - AM)} d M
    \end{equation}
with a general potential $V$. The Riemann-Hilbert problem is formulated
in Section 2.

The asymptotic analysis of the Riemann-Hilbert problem presents a
new feature that we feel is of importance in its own right. We will not use
the Pastur equation (\ref{PasturEquation}) which defines the $\xi$-functions
and the Riemann surface that corresponds to it,
but instead we use a modified equation to define the $\xi$-functions.
We discuss this in Section 3. The modification may be thought of in
potential theoretic terms and we briefly discuss this in Section 3 as well.

The anti-derivatives of the modified $\xi$-functions are introduced in
Section 4 and they play an important role in the steepest descent analysis of
the Riemann-Hilbert problem in the rest of the paper. The main issue is
the construction in Section 8 of the local parametrix around $0$ with the aid
of Pearcey integrals. The modification of the $\xi$-functions is used here
to be able to match the local Pearcey parametrix with the outside
parametrix. Even so it turns out that we cannot achieve the matching condition
on a fixed circle around the origin, but only on circles with radii $n^{-1/4}$ that
decrease as $n$ increases. However, the circles are big enough to capture
the behavior (\ref{limitkernel}) which takes place at a distance to the origin
of order $n^{-3/4}$. The precise estimates that lead to the proof of
Theorem \ref{maintheo} are given in the final Sections 9 and 10.

\section{Riemann-Hilbert problem}
As shown in our paper \cite{BK1}, the correlation kernel is expressed in terms of
the solution to the
following $3 \times 3$ matrix valued Riemann-Hilbert (RH) problem.

Find $Y:\C\setminus\R\to\C^{3\times 3}$ such that
\begin{itemize}
\item $Y$ is analytic on $\C\setminus\R$,
\item for $x\in\R$, we have
\begin{equation}\label{m4}
Y_+(x)=Y_-(x)
\begin{pmatrix}
1 & e^{-n(\frac{1}{2} x^2-ax)} & e^{-n(\frac{1}{2} x^2+ax)} \\
0 & 1 & 0 \\
0 & 0 & 1
\end{pmatrix},
\end{equation}
where $Y_+(x)$ ($Y_-(x)$) denotes the limit of $Y(z)$ as $z\to x$ from
the upper (lower) half-plane,
\item as $z\to\infty$, we have
\begin{equation}\label{m6}
Y(z)=\left( I+O\left(\frac{1}{z}\right)\right)
\begin{pmatrix}
z^{n} & 0 & 0 \\
0 & z^{-n/2} & 0 \\
0 & 0 & z^{-n/2}
\end{pmatrix},
\end{equation}
where $I$ denotes the $3\times 3$ identity matrix.
\end{itemize}

The RH problem has a unique solution,  given explicitly in terms
of the multiple Hermite polynomials. The correlation
kernel of the Gaussian random matrix model with external source
is equal to
\begin{equation}
    \label{corrkernel} K_n(x,y;a) =
    \frac{e^{-\frac{1}{4}n(x^2+y^2)}}{2\pi i (x-y)}
    \begin{pmatrix} 0 & e^{nay} & e^{-nay} \end{pmatrix}
    Y^{-1}_+(y) Y_+(x)
    \begin{pmatrix} 1 \\ 0 \\ 0 \end{pmatrix}.
\end{equation}

In what follows we are going to apply the Deift/Zhou steepest
descent method for RH problems to the above RH problem for $Y$.
It consists of a sequence of explicit transformations
$Y \mapsto T \mapsto S \mapsto R$ which leads to a RH
problem for $R$ in which all jumps are close to the identity
matrix and which is normalized at infinity. Then $R$ is close
to the identity matrix, and analyzing the effect of the
transformations on the kernel (\ref{corrkernel}) we will be
able to prove Theorem \ref{maintheo}.

\section{Modification of the $\xi$-functions}

\subsection{Modified Pastur equation}
The analysis in \cite{ABK,BK2} for the cases $a > 1$ and $0 < a < 1$
was based on the equation (\ref{PasturEquation}) and it would be natural
 to use (\ref{PasturEquation}) also in the case $a=1$.
Indeed, that is what we tried to do, and we found that it works for $a\equiv 1$, but
in the double scaling regime $a= 1 + \frac{b}{2\sqrt{n}}$
with $b \neq 0$, it led to problems that we were unable to resolve
in a satisfactory way.
A crucial feature of our present approach is a modification of
the equation (\ref{PasturEquation}) when $a$ is close to $1$,
but different from $1$. At $x=0$ we wish to have a double
branch point for all values of $a$ so that the structure
of the Riemann surface is as in the middle figure of Figure \ref{fig1}
for all $a$.

For $c > 0$, we consider the Riemann surface for the equation
\begin{equation} \label{RSeq}
    z = \frac{w^3}{w^2 -c^2}
\end{equation}
where $w$ is a new auxiliary variable.
The Riemann surface has branch points at $z^* = \frac{3\sqrt{3}}{2} c$,
$-z^*$ and a double branch point at $0$.
There are three inverse functions $w_k$, $k=1,2,3$, that
behave as $z \to \infty$ as
\begin{equation} \label{wasym}
    \begin{aligned}
    w_1(z) &= z - \frac{c^2}{z} + O\left(\frac{1}{z^3}\right) \\
    w_2(z) &= c + \frac{c^2}{2z} + O\left(\frac{1}{z^2}\right) \\
    w_3(z) &= -c + \frac{c^2}{2z} + O\left(\frac{1}{z^2}\right)
    \end{aligned} \end{equation}
and which are defined and analytic on $\mathbb C \setminus [-z^*,z^*]$,
$\mathbb C \setminus [0, z^*]$ and $\mathbb C \setminus [-z^*,0]$,
respectively.

Then we define the modified $\xi$-functions
\begin{equation} \label{defxik}
    \xi_k = w_k + \frac{p}{w_k}, \qquad \mbox{for } k=1,2,3,
\end{equation}
which we also consider on their respective Riemann sheets.
In what follows we take
\begin{equation} \label{defc}
    c = \frac{a + \sqrt{a^2+8}}{4} \quad
        \mbox{ and } \quad p = c^2-1.
\end{equation}
Note that $a = 1$ corresponds to $c=1$ and $p=0$. In that case
the functions coincide with the solutions of the equation
(\ref{PasturEquation}) that we used in our earlier works.
From (\ref{RSeq}), (\ref{defxik}), and (\ref{defc}) we obtain
the modified Pastur equation
\begin{equation} \label{modPE}
    \xi^3 - z \xi^2  + (1-a^2) \xi + a^2 z
        + \frac{(c^2-1)^3}{c^2 z} = 0,
\end{equation}
where $c$ is given by (\ref{defc}).

\begin{lem}
Let $a > 0$ and take $c$ and $p$ as in {\rm (\ref{defc})}.
Then at infinity we have
\begin{equation}\label{sch6}
\begin{aligned}
\xi_1(z)&=z-\frac{1}{z}+O\left(\frac{1}{z^3}\right),\\
\xi_2(z)&=a+\frac{1}{2z}+O\left(\frac{1}{z^2}\right),\\
\xi_3(z)&=-a+\frac{1}{2z}+O\left(\frac{1}{z^2}\right).
\end{aligned}
\end{equation}
\end{lem}
\begin{proof}
This follows from direct calculations using
(\ref{wasym}), (\ref{defxik}) and the fact that $2c - \frac{1}{c} = a$.
Alternatively, one could also use (\ref{modPE}).
\end{proof}

The new $\xi$-functions have the same asymptotic behavior (\ref{sch6}) as $z \to \infty$
(up to order $1/z^2$)
as the solutions of (\ref{PasturEquation}). This is important for
the first two transformations of the Riemann-Hilbert problem.
The situation at $z=0$ is different. The fact that we can
control the behavior at $z=0$ as well is the reason for the introduction
of the modified $\xi$-functions.

\subsection{Behavior at $z=0$}
We start with the behavior of the functions $w_k$.
\begin{lem} \label{lemwkat0}
There exist analytic functions $f_1$ and $g_1$
defined in a neighborhood $U_1$ of $z=0$ so that
for $z \in U_1$ and $k=1,2,3$,
\begin{equation} \label{wkf1g1}
    w_k(z) =
    \left\{ \begin{array}{ll} \ds
     - \omega^{2k} z^{1/3}  f_1(z) -
    \omega^k z^{5/3} g_1(z) + \frac{z}{3}
        & \mbox{ for } \Im z > 0, \\[10pt] \ds
     - \omega^k z^{1/3}  f_1(z) -
    \omega^{2k} z^{5/3} g_1(z) + \frac{z}{3}
        & \mbox{ for } \Im z < 0.
    \end{array} \right.
    \end{equation}
In addition, we have $f_1(0) = c^{2/3}$, and $f_1(z)$ and $g_1(z)$
are real for real $z \in U_1$.
\end{lem}

\begin{proof}
Putting $z = x^3$ and $w = xy$ in (\ref{RSeq}) we obtain
\begin{equation} \label{RSeq2}
    y^3 = x^2 y^2 - c^2
\end{equation} which has a solution $y=y(x)$
that is analytic in a neighborhood $U_1$ of $0$ and
satisfies $y(0) = -c^{2/3}$ and $y'(0) = 0$.
Then we can write $y(x) = -f_1(x^3) -x^4 g_1(x^3) + x^2 h_1(x^3)$
with $f_1$, $g_1$ and $h_1$ analytic in $U_1$ and
$f_1(0) = c^{2/3}$.
Putting this back in (\ref{RSeq2}) we find after straightforward
calculations that (with $z=x^3$)
\begin{equation} \label{f1g1A}
    f_1(z) g_1(z) = \frac{1}{9}, \quad
    f_1(z)^3 - c^2 +  \frac{2}{27} z^2 +  g_1(z)^3 z^4  = 0,
\end{equation}
and $h_1(z) = \frac{1}{3}$.
Going back to $z$ and $w$ variables, we see that there is a solution
$w = w(z)$ to (\ref{RSeq}) with
\[ w(z) = - z^{1/3} f_1(z) - z^{5/3} g_1(z) + \frac{z}{3},
    \qquad \mbox{ for } z \in \mathbb C \setminus (-\infty,0], \]
where we take the principal branches of the fractional powers.
This solution is real for $z$ real and positive, and so it coincides
with the solution $w_3(z)$. This proves (\ref{wkf1g1}) for $k=3$.
The expressions (\ref{wkf1g1}) for $k=1,2$ follow by analytic
continuation.

Since $y(x)$ is real for real $x$, we also find that $f_1(z)$
and $g_1(z)$ are real if $z$ is real.
\end{proof}

From (\ref{f1g1A}) it is easy to give explicit
expressions for $f_1$ and $g_1$. However we will not use this
in the future.

From Lemma \ref{lemwkat0} and (\ref{defxik})
we get the following behavior for the functions $\xi_k$ near $z=0$.

\begin{lem} \label{lemxikat0}
There exist analytic functions $f_2$ and $g_2$
defined in a neighborhood $U_2$ of $z=0$ so that
for $z \in U_2$ and $k=1,2,3$,
\begin{equation} \label{xikf2g2} \xi_k(z) =
    \left\{ \begin{array}{ll} \ds
     - \omega^{2k} z^{1/3}  f_2(z) -
    \omega^k z^{-1/3} g_2(z) + \frac{z}{3}
        & \mbox{ for } \Im z > 0, \\[10pt] \ds
     - \omega^k z^{1/3}  f_2(z) -
    \omega^{2k} z^{-1/3} g_2(z) + \frac{z}{3}
        & \mbox{ for } \Im z < 0.
    \end{array} \right. \end{equation}
In addition, we have
\begin{equation} \label{f2g2at0}
    f_2(0) = c^{2/3} + \frac{1}{3}c^{-4/3}(c^2-1),
    \qquad
    g_2(0) = c^{-2/3}(c^2-1),
\end{equation}
and $f_2(z)$ and $g_2(z)$ are real for real $z \in U_2$.
\end{lem}

\begin{proof}
This follows from (\ref{defxik}) and the previous lemma. Indeed from (\ref{wkf1g1})
where  $f_1$ and $g_1$ satisfy the relations
(\ref{f1g1A}), we can deduce
\[ \frac{1}{w_3(z)} =
    - z^{1/3} \frac{1}{3c^2} (f_1(z) + 3 z^2 g_1(z)^2)
    - z^{-1/3} \frac{1}{3c^2} (3 f_1(z)^2 + z^2 g_1(z)).
    \]
Then (\ref{xikf2g2}) follows from (\ref{wkf1g1}) and
(\ref{defxik}) if we take
\begin{equation} \label{deff2}
    f_2(z) = f_1(z) + \frac{c^2-1}{3c^2} (f_1(z) + 3z^2 g_1(z)^2)
\end{equation}
and
\begin{equation} \label{defg2}
    g_2(z) = z^2 g_1(z) + \frac{c^2-1}{3c^2}(3f_1(z)^2 + z^2 g_1(z)).
\end{equation}
Because of Lemma \ref{lemwkat0} this also implies (\ref{f2g2at0}) and the fact that $f_2(z)$
and $g_2(z)$ are real for real $z \in U_2$.
\end{proof}

\subsection{Potential theoretic interpretation}
As an aside we want to mention that the modified $\xi$-functions may be thought
of in terms of a modified equilibrium problem for logarithmic potentials.
For $a > 1$, it was noted in \cite{BK2}, that the limiting mean eigenvalue
density $\rho(x) = \rho(x;a)$ may be characterized as follows. We minimize
\begin{equation} \label{minE}
    \begin{aligned}
        E(\mu_1,\mu_2) = & \iint \log \frac{1}{|x-y|} d\mu_1(x) d\mu_1(y)
        + \iint \log \frac{1}{|x-y|} d\mu_2(x) d\mu_2(y) \\
        & + \iint \log \frac{1}{|x-y|} d\mu_1(x) d\mu_2(y) +
        \int \left( \frac{1}{2} x^2 - ax \right) d\mu_1(x) \\
        & + \int \left( \frac{1}{2} x^2 + ax \right) d\mu_2(x)
        \end{aligned}
        \end{equation}
among all non-negative measures $\mu_1,\mu_2$ on $\mathbb R$
with $\int d\mu_1 = \int d\mu_2 = \frac{1}{2}$. There is a unique minimizer
\cite{ST}, and for $a > 1$, we have that $\supp(\mu_1) \subset [0,\infty)$,
$\supp(\mu_2) \subset (-\infty,0]$, and $\rho$ is the density of $\mu_1 + \mu_2$.
For $a < 1$, the minimizing measures for (\ref{minE}) do not have disjoint
supports, and in fact these minimizers are not related to our random matrix
ensemble (\ref{matrixmodel}) at all.

The modification we are alluding to is to minimize (\ref{minE}) among
{\em signed} measures $\mu_1 = \mu_1^+ - \mu_1^-$, $\mu_2 = \mu_2^+ - \mu_2^-$
where $\mu_j^{\pm}$ are non-negative measures, that satisfy
$\int d\mu_1 = \int d\mu_2 = \frac{1}{2}$ and in addition
\begin{itemize}
\item[(1)] $\supp(\mu_1) \subset [0, \infty)$, $\supp(\mu_2) \subset (-\infty, 0]$, and,
\item[(2)] There is a $\delta > 0$, such that $\supp(\mu_1^-) \subset [0,\delta)$
and $\supp(\mu_2^-) \subset (-\delta, 0]$.
\end{itemize}
The condition (1) plays a role for $a < 1$, since it prevents the supports of
$\mu_1$ and $\mu_2$ to overlap. For $a > 1$, the condition (2) plays a role, since
it allows the measures to become negative near $0$.
Now let $\mu_1, \mu_2$ be the minimizers for this modified equilibrium problem,
and let $\tilde{\rho}$ be the density of $\mu_1 + \mu_2$. Then it can
be shown that the density of $\mu_1 + \mu_2$ is equal to
$\frac{1}{\pi} \Im \xi_{1+}(x)$ where $\xi_1$ is the modified
$\xi_1$-function introduced in this section.

We will not use this potential-theoretic connection in the
analysis that follows in this paper, but we anticipate that it might be
important for the general unitary random matrix ensemble with
external source (\ref{generalmatrixmodel}).

We finally note that a modified equilibrium problem was also used
in \cite{CK,CKV} in order to analyse the double scaling limit in
unitary random matrix ensembles (without external source), so one might
speculate that such an approach might be characteristic for
double scaling limits in random matrix ensembles.

\section{The $\lambda$-functions}

\subsection{Definition and first properties}
The main role is played by the $\lambda$-functions which
 are anti-derivatives of the $\xi$-functions. They are defined here as
\begin{equation} \label{deflambdak}
    \lambda_k(z) = \int_{0+}^z \xi_k(s) ds
\end{equation}
where the path of integration starts at $0$ on the upper
side of the cut and is fully contained (except for the initial point)
in $\mathbb C \setminus (-\infty,z^*]$
for $k=1,2$, and in $\mathbb C \setminus (-\infty,0]$ for $k=3$.
Then $\lambda_1$ and $\lambda_2$ are defined and analytic
on $\mathbb C \setminus (-\infty,z^*]$, and $\lambda_3$
is defined and analytic on $\mathbb C \setminus (-\infty,0]$.

As follows from (\ref{sch6}) and (\ref{deflambdak}),
the $\la$-functions behave at infinity as
\begin{equation} \label{lambdaatinfinity}
\begin{aligned}
    \lambda_1(z) &= \frac{1}{2}z^2 - \log z + \ell_1 + O(1/z), \\
    \lambda_2(z) &= az + \frac{1}{2} \log z + \ell_2 + O(1/z), \\
    \lambda_3(z) &= -az + \frac{1}{2} \log z + \ell_3 + O(1/z),
\end{aligned}
\end{equation}
for certain constants $\ell_k$, $k=1,2,3$, where $\log z$
is taken as the principal value, that is, with a cut along the negative
real axis.

From contour integration based on (\ref{sch6}) where we use the
residue of $\xi_2$ at infinity, we find $\lambda_{1-}(0) = \pi i$
and $\lambda_{2-}(0) = -\pi i$. Then we get the following
jump properties of the $\lambda$-functions on the cuts
$(-\infty,0]$ and $(-\infty,z^*]$:
\begin{equation} \label{lambdajumps}
\begin{array}{llll}
    \lambda_{1+} = \lambda_{2-} + \pi i,
    & \lambda_{2+} = \lambda_{1-} - \pi i,
    & \lambda_{3+} = \lambda_{3-}
    \quad & \mbox{ on } [0, z^*],  \\[5pt]
    \lambda_{1+} = \lambda_{3-}, \quad
    & \lambda_{2+} = \lambda_{2-} + \pi i,
    & \lambda_{3+} = \lambda_{1-} -\pi i,
    \quad & \mbox{ on } [-z^*,0],  \\[5pt]
    \lambda_{1+} = \lambda_{1-} - 2\pi i,
    & \lambda_{2+} = \lambda_{2-} + \pi i,
    & \lambda_{3+} = \lambda_{3-} + \pi i,
    \quad & \mbox{ on } (-\infty,-z^*].
\end{array}
\end{equation}

\subsection{Behavior near $z=0$}
Near the origin the $\lambda$-functions behave as follows.
\begin{lem} \label{lambdakat0}
There exist analytic functions $f_3$ and $g_3$ in
a neighborhood $U_3$ of $z=0$ so that
\begin{equation} \label{lambdakf3g3}
    \lambda_k(z) =
    \left\{ \begin{array}{ll} \ds
     - \frac{3}{4} \omega^{2k} z^{4/3}  f_3(z)
     - \frac{1}{2} \omega^k z^{2/3} g_3(z) + \frac{z^2}{6}
        & \mbox{ for } \Im z > 0, \\[10pt] \ds
    \lambda_{k-}(0) - \frac{3}{4} \omega^k z^{4/3}  f_3(z)
    - \frac{1}{2}  \omega^{2k} z^{2/3} g_3(z) + \frac{z^2}{6}
        & \mbox{ for } \Im z < 0,
    \end{array} \right. \end{equation}
In addition, we have
\begin{equation} \label{f3g3at0}
f_3(0) = f_2(0) = c^{2/3} + \frac{1}{3}c^{-4/3}(c^2-1),
\quad
g_3(0) = 3 g_2(0) = 3 c^{-2/3}(c^2-1),
\end{equation}
and $f_3(z)$ and $g_3(z)$ are real for real $z \in U_3$.
\end{lem}
\begin{proof}
The relations (\ref{lambdakf3g3}) follow by
integrating (\ref{xikf2g2}). Note that $\lambda_{1-}(0) = \pi i$,
$\lambda_{2-}(0) = -\pi i$, and $\lambda_{3-}(0)= 0$.
The other statements of the lemma also follow directly from Lemma \ref{lemxikat0}.
\end{proof}

\subsection{Critical trajectories}
\begin{figure}[t]
\centerline{\includegraphics[height=7cm,width=12cm]{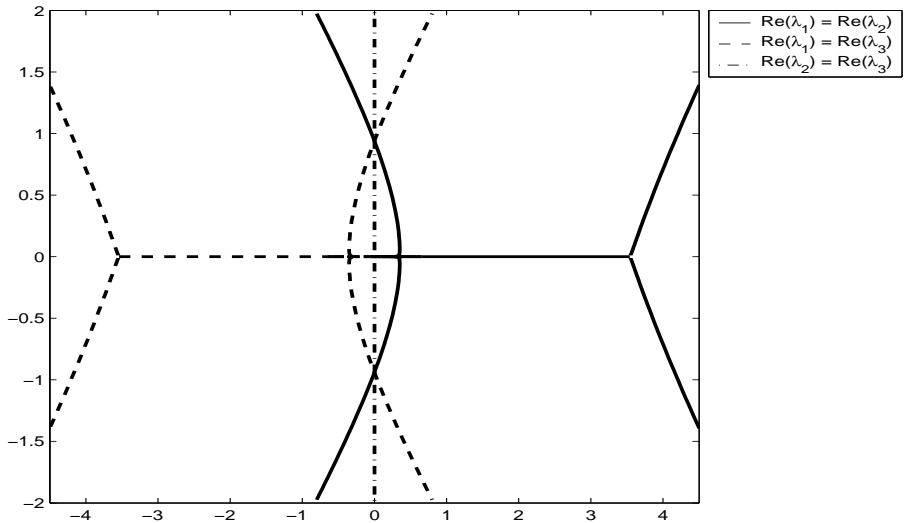}}
\caption{Curves where $\Re \lambda_j = \Re \lambda_k$ for
the value $a=2.0$.}
\label{fig:curves20}
\end{figure}

\begin{figure}[t]
\centerline{\includegraphics[height=7cm,width=12cm]{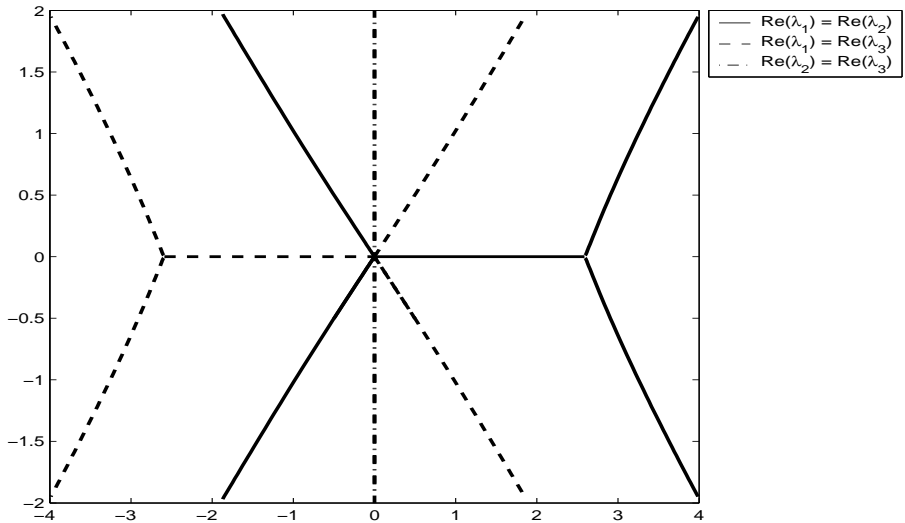}}
\caption{Curves where $\Re \lambda_j = \Re \lambda_k$ for
the value $a = 1.0$.}
\label{fig:curves10}
\end{figure}

\begin{figure}[t]
\centerline{\includegraphics[height=7cm,width=12cm]{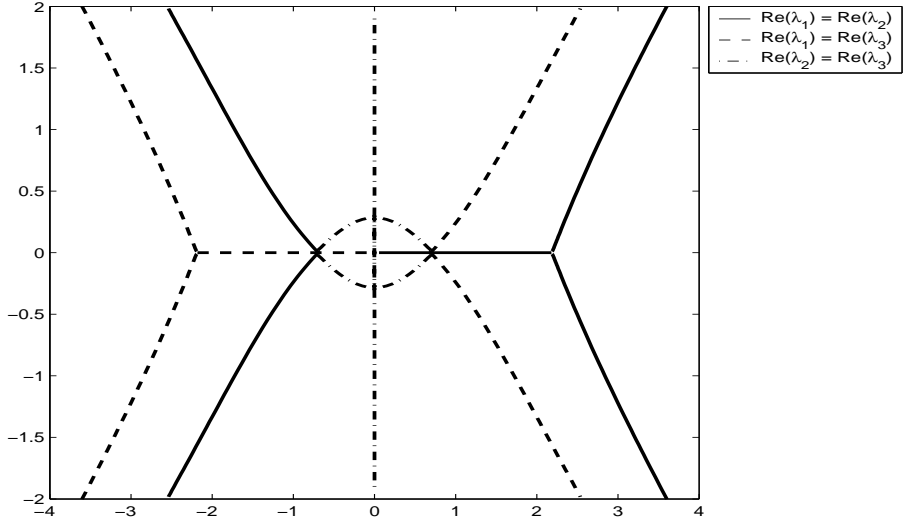}}
\caption{Curves where $\Re \lambda_j = \Re \lambda_k$ for
the value $a = 0.5$.}
\label{fig:curves05}
\end{figure}

Curves where $\Re \lambda_j  = \Re \lambda_k$ for some $j \neq k$
are shown in Figures \ref{fig:curves20}, \ref{fig:curves10}, and \ref{fig:curves05},
for the cases $a > 1$, $a=1$, and $a < 1$, respectively.
These are critical curves that play a crucial role in the asymptotic analysis.
The curves are
critical trajectories of the quadratic differentials
$(\xi_j(z)-\xi_k(z))^2 dz^2$ (and their analytic continuations
beyond the branch cuts in case $a < 1$).

The solid curves in Figures \ref{fig:curves20}--\ref{fig:curves05} are
the critical trajectories of the quadratic differential
$(\xi_1(z) - \xi_2(z))^2 dz^2$. The quadratic differential has a simple zero
at $z = z^*$. Three trajectories are emanating from
$z = z^*$ at equal angles, one of these being the real interval $(0,z^*)$.
For $a > 1$, the quadratic differential has a double zero at $z = x_0$ for
some $x_0 \in (0, z^*)$. Four trajectories
are emanating from the double zero at equal angles as can be seen
in Figure \ref{fig:curves20}.

The dashed curves are the critical trajectories of the
quadratic differential $(\xi_1(z) - \xi_3(z))^2 dz^2$.
Because of symmetry, these are the mirror images of the trajectories of the
quadratic differential $(\xi_1(z) - \xi_2(z))^2 dz^2$ with
respect to  the imaginary axis. For $a > 1$, the solid curve
that passes vertically through $x_0$ and its dashed mirror image with
respect to the imaginary axis
meet in two points $\pm iy_0$ on the imaginary axis.
Together they enclose a neighborhood of the origin.

The dashed-dotted curves are the critical trajectories of
$(\xi_2(z)-\xi_3(z))^2 dz^2$. For $a < 1$, the quadratric differential
has two double zeros at $z= \pm iy_0$ for some  $y_0 > 0$. Four
trajectories are emanating from these double zeros at equal
angles as shown in the Figure \ref{fig:curves05}. Besides the imaginary
axis there are curves passing horizontally through $\pm iy_0$,
and these curves meet each other at two points $\pm x_0$
on the real axis and they enclose a neighborhood of the origin.
Beyond these two points the quadratic differentials have analytic
continuations, but the formula changes since either $\xi_2$ or $\xi_3$ reaches its
branch cut and changes into $\xi_1$. Consequently, the dashed-dotted curves
in Figure \ref{fig:curves05} continue beyond $\pm x_0$ as either solid or dashed curves.

The relative orderings of the real parts  $\Re \lambda_1$, $\Re \lambda_2$ and $\Re \lambda_3$
changes if we cross one of the critical trajectories,
but it remains constant in the regions bounded by the critical trajectories.
For each of the unbounded regions we can determine
the ordering from the behavior at infinity
(\ref{lambdaatinfinity}). For example, we have in the right-most
region $\Re \lambda_1 > \Re \lambda_2 > \Re \lambda_3$, and
if we cross the solid curve where $\Re \lambda_1 = \Re \lambda_2$,
the ordering becomes $\Re \lambda_2 > \Re \lambda_1 > \Re \lambda_3$,
and so on.

In the cases $a < 1$ and $a > 1$ the trajectories enclose a bounded
neighborhood of the origin. There is
no such neighborhood in case $a = 1$. The neighborhood
is small if $a$ is close to $1$. In this neighborhood
the relative ordering of the real parts is different.

So we can easily verify the following.

\begin{lem} \label{ordering}
Except for $z$ in the exceptional bounded neighborhood
of the origin, we have that
\[ \Re \lambda_2(z) > \max(\Re \lambda_1(z), \Re \lambda_3(z)) \]
in the region in the right-half plane, bounded by the solid
and dashed-dotted curves, and
\[ \Re \lambda_3(z) > \max(\Re \lambda_1(z), \Re \lambda_2(z)) \]
in the region in the left-half plane bounded by the dashed and
dashed-dotted curves.
\end{lem}

The exceptional neighborhood will not cause a problem to us,
since it turns out to shrink fast enough if $a = 1 + (b/2)n^{-1/2}$
and $n \to \infty$.
For $n$ large enough, the exceptional neighborhood is well
within the disk around the origin of radius $n^{-1/4}$ where
we are going to construct a special parametrix with Pearcey integrals.
Then the different ordering of the real parts of the $\lambda_k$
will not play a role.

\section{First two transformations of the RH problem}
\label{sec3}
The first and second transformation of the RH problem are
the same as in our earlier paper \cite{BK2}, except that we use
the $\lambda$-functions that were introduced in the last section
via the modified $\xi$-functions.

\subsection{First transformation $Y \mapsto T$} \label{subsec31}
Using the functions $\lambda_k$ and the constants $\ell_k$, $k=1,2,3$,
we define
\begin{equation} \label{defT}
   T(z) = \diag\left(e^{-n\ell_1}, e^{-n\ell_2}, e^{-n\ell_3}\right)
    Y(z) \diag\left(e^{n(\la_1(z)-\frac{1}{2}z^2)}, e^{n(\la_2(z) - az)}, e^{n(\la_3(z)+az)} \right).
\end{equation}
Then by (\ref{m4}) and (\ref{defT}) and the jump properties
(\ref{lambdajumps}) we have $T_+(x) = T_-(x) j_T(x)$ for
$x \in \mathbb R$, where
\begin{equation}\label{jT1}
\begin{aligned}
j_T &=\begin{pmatrix}
e^{n(\la_1-\la_2)_+} & 1 & e^{n(\la_{3} - \la_{1-})} \\
0 & e^{n(\la_1-\la_2)_-} & 0 \\
0 & 0 & 1
\end{pmatrix},
\end{aligned}
\qquad x \in (0,z^*),
\end{equation}
\begin{equation}\label{jT2}
\begin{aligned}
j_T&=\begin{pmatrix}
e^{n(\la_1-\la_3)_+} & e^{n(\la_{2+}-\la_{1-})} & 1 \\
0 & 1 & 0 \\
0 & 0 & e^{n(\la_1-\la_3)_-}
\end{pmatrix},
\end{aligned}
\qquad
x \in (-z^*, 0),
\end{equation}
\begin{equation}\label{jT3}
j_T=
\begin{pmatrix}
1 & e^{n\left(\la_{2+}-\la_{1-}\right)}
& e^{n\left(\la_{3+}-\la_{1-}\right)} \\
0 & 1 & 0 \\
0 & 0 & 1
\end{pmatrix},\quad
x \in (-\infty,-z^*) \cup (z^*, \infty).
\end{equation}
The function $T(z)$ solves the following RH problem:
\begin{itemize}
\item $T$ is analytic on $\C\setminus\mathbb R$,
\item for $x \in \mathbb R$, we have
\begin{equation}\label{ft14}
T_+(x)=T_-(x)j_T(x),
\end{equation}
where $j_T$ is given by (\ref{jT1})-(\ref{jT3}),
\item as $z\to\infty$,
\begin{equation}\label{ft15}
T(z)=I+O\left(\frac{1}{z}\right).
\end{equation}
\end{itemize}
The asymptotic property (\ref{ft15}) follows from (\ref{m6}),
(\ref{defT}),
and the behavior (\ref{lambdaatinfinity}) of the $\lambda$-functions
at infinity.

\subsection{Second transformation $T \mapsto S$}
\label{subsec32}

\begin{figure}[t]
\centerline{\includegraphics[height=7cm,width=12cm]{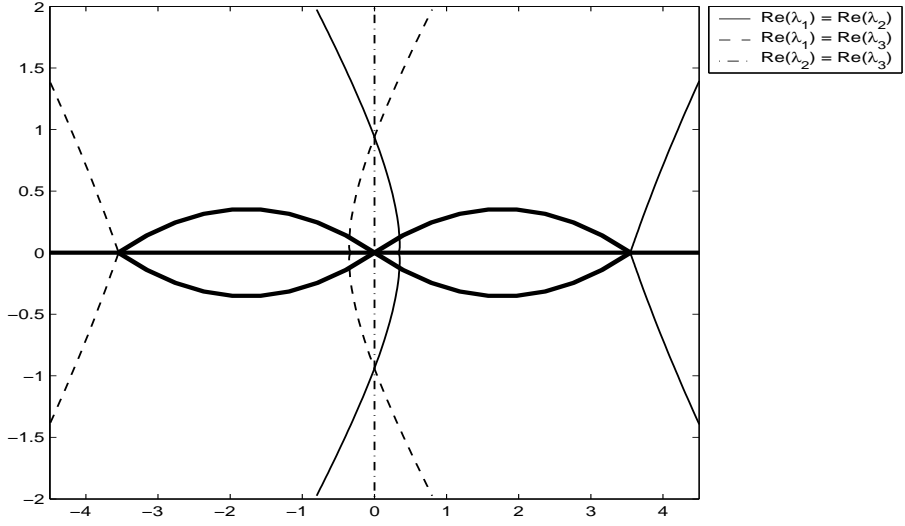}}
\caption{Opening of lenses around the intervals $[0,z^*]$ and $[-z^*,0]$ for the value $a=2.0$.
The upper and lower lips of the lenses together with the real axis form the contour $\Sigma_S$
which is shown in bold.
Also shown are the critical trajectories  as in Figure \ref{fig:curves20}.}
\label{fig:opening20}
\end{figure}

The second transformation of the RH problem consists of opening of
lenses around the intervals $[0,z^*]$ and $[-z^*,0]$.
The lenses are as shown in Figure \ref{fig:opening20}.
We define (see also Section 5 in \cite{BK2})
\begin{align} \label{defS1}
    S = T \begin{pmatrix} 1&0&0 \\
    -e^{n(\la_1-\la_2)}&1& -e^{n(\la_3-\la_2)}\\
    0&0&1 \end{pmatrix}
    & \qquad \mbox{ in the upper right lens region,} \\
    \label{defS2}
    S = T \begin{pmatrix} 1&0&0 \\
    e^{n(\la_1-\la_2)}&1& -e^{n(\la_3-\la_2)}\\
    0&0&1 \end{pmatrix}
    & \qquad \mbox{ in the lower right lens region,}  \\
    \label{defS3}
    S = T \begin{pmatrix} 1&0&0 \\
    0&1&0\\
    -e^{n(\la_1-\la_3)}&-e^{n(\la_2-\la_3)} &1 \end{pmatrix}
    & \qquad \mbox{ in the upper left lens region,} \\
    \label{defS4}
    S = T \begin{pmatrix} 1&0&0 \\ 0&1&0 \\
    e^{n(\la_1-\la_3)}& -e^{n(\la_2-\la_3)} & 1 \end{pmatrix}
    & \qquad \mbox{ in the lower left lens region,}
    \end{align}
and $S = T$ outside the lenses.

It leads to a matrix valued function $S$
which is defined and analytic in $\mathbb C \setminus \Sigma_S$,
where $\Sigma_S$ consists of the real line and the upper and lower lips
of the lenses. On $\Sigma_S$ we have $S_+ = S_- j_S$ where
$j_S$ is defined as follows (the orientation on $\Sigma_S$
is taken from left to right):
\begin{align} \label{jS1}
j_S =  \begin{pmatrix}
0 & 1 & 0 \\
-1& 0 & 0 \\
0 & 0 & 1
\end{pmatrix},
& \quad \textrm{ on } (0,z^*),\\ \label{jS2}
j_S = \begin{pmatrix}
0 & 0 & 1 \\
0 & 1 & 0 \\
-1& 0 & 0
\end{pmatrix},
& \quad \textrm{ on } (-z^*,0),\\ \label{jS3}
j_S = \begin{pmatrix}
1 & e^{n(\lambda_{2+} - \lambda_{1-})} &
    e^{n(\lambda_{3+} - \lambda_{1-})} \\
0 & 1 & 0 \\
0 & 0 & 1
\end{pmatrix},
& \quad \textrm{ on } (-\infty,-z^*) \cup (z^*,\infty),\\ \label{jS4}
j_S = \begin{pmatrix}
1 & 0 & 0 \\
e^{n(\lambda_1-\lambda_2)} & 1 & e^{n(\lambda_3-\lambda_2)} \\
0 & 0 & 1
\end{pmatrix},
& \quad \textrm{ on the upper lip of the right lens},\\ \label{jS5}
j_S = \begin{pmatrix}
1 & 0 & 0 \\
0 & 1 & 0 \\
e^{n(\lambda_1-\lambda_3)} & e^{n(\lambda_2-\lambda_3)} & 1
\end{pmatrix},
& \quad \textrm{ on the upper lip of the left lens},\\ \label{jS6}
j_S = \begin{pmatrix}
1 & 0 & 0 \\
0 & 1 & 0 \\
e^{n(\lambda_1-\lambda_3)} & -e^{n(\lambda_2-\lambda_3)} & 1
\end{pmatrix},
& \quad \textrm{ on the lower lip of the left lens},\\ \label{jS7}
j_S = \begin{pmatrix}
1 & 0 & 0 \\
e^{n(\lambda_1-\lambda_2)} & 1 & -e^{n(\lambda_3-\lambda_2)} \\
0 & 0 & 1
\end{pmatrix},
& \quad \textrm{ on the lower lip of the right lens}.
\end{align}
Thus $S$ solves the following RH problem:
\begin{itemize}
\item $S$ is analytic on $\C\setminus\Sigma_S$,
\item for $z \in \Sigma_S$, we have $S_+(z)=S_-(z)j_S(z)$,
where $j_S$ is given by (\ref{jS1})-(\ref{jS7}),
\item as $z\to\infty$, we have $S(z)=I+O\left(\frac{1}{z}\right)$.
\end{itemize}

Now the ordering of the real parts of the $\lambda_k$
in various regions in the complex plane (see Lemma \ref{ordering})
shows that the jump matrices in (\ref{jS3})--(\ref{jS7})
are all close to the identity matrix if $n$ is large,
except in a neighborhood of the origin. To be precise, if $a < 1$ then
$\Re \lambda_3 > \Re \lambda_2$ near the origin in
the right half-plane, which means that the entries $\pm e^{n(\lambda_3-\lambda_2)}$
in the jump matrices in (\ref{jS4}) and (\ref{jS7}) are not small near
the origin but instead grow exponentially  if $n$ gets large.
Similarly the entries $\pm e^{n(\lambda_2-\lambda_3)}$
in the jump matrices in (\ref{jS5}) and (\ref{jS6}) also grow exponentially
near the origin.
On the other hand, if $a > 1$, then $\Re \lambda_1$
is bigger than the other two in the exceptional neighborhood of the origin,
so that the other non-zero off-diagonal entries
in the jump matrices in (\ref{jS4})-(\ref{jS7}) grow exponentially in
a neighborhood of the origin.
For $a =1$ there are no such exceptions and all jump matrices in
(\ref{jS3})-(\ref{jS7}) are close to the identity matrix if $n$ is large.

When we wrote that certain entries grow exponentially as $n$ gets large,
it was understood that the value of $a \neq 1$ remained fixed. However,
eventually we are going to take $a = 1 + O(n^{-1/2})$ as $n \to \infty$.
Then it will turn out that the possible growth of certain entries in
the jump matrices is confined to a small enough region near the origin, which
shrinks sufficiently fast as $n \to \infty$, so that
we can still ignore the jumps (\ref{jS3})-(\ref{jS7})
in the next step.

\section{Model RH Problem}

We consider the following auxiliary model RH problem: find $M:
\C \setminus [-z^*,z^*] \to \C^{3\times 3}$ such that
\begin{itemize}
\item $M$ is analytic on $\C\setminus [-z^*,z^*]$,
\item for $x \in(-z^*,z^*)$ we have $M_+(x)=M_-(x)j_M(x)$,
where
\begin{equation}\label{mod2}
j_M(x) =
\begin{pmatrix}
0 & 1 & 0 \\
-1 & 0 & 0 \\
0 & 0 & 1
\end{pmatrix},\qquad \textrm{ for } x\in (0,z^*),
\end{equation}
and
\begin{equation}\label{mod3}
j_M(x)=
\begin{pmatrix}
0 & 0 & 1 \\
0 & 1 & 0 \\
-1 & 0 & 0
\end{pmatrix},\qquad \textrm{ for } x\in (-z^*,0),
\end{equation}
\item as $z\to\infty$,
\begin{equation}\label{mod4}
M(z)=I+O\left(\frac{1}{z}\right).
\end{equation}
\end{itemize}
This RH problem has a solution, see \cite[Section 6]{BK2},
that can be explicitly given in terms of the mapping functions
$w_k$, $k=1,2,3$, from (\ref{RSeq}) and (\ref{wasym}).
The solution takes the form
\begin{equation} \label{defM}
    M(z) = \begin{pmatrix}
    M_1(w_1(z)) & M_1(w_2(z)) & M_1(w_3(z)) \\
    M_2(w_1(z)) & M_2(w_2(z)) & M_2(w_3(z)) \\
    M_3(w_1(z)) & M_3(w_2(z)) & M_3(w_3(z))
    \end{pmatrix} \end{equation}
where $M_1$, $M_2$, $M_3$ are the three scalar valued functions
\begin{equation} \label{defM123}
    M_1(w) = \frac{w^2-c^2}{w \sqrt{w^2-3c^2}}, \quad
    M_{2}(w) = \frac{-i}{\sqrt{2}} \frac{w +c}{w \sqrt{w^2-3c^2}}, \quad
    M_{3}(w) = \frac{-i}{\sqrt{2}} \frac{w -c}{w \sqrt{w^2-3c^2}}.
    \end{equation}

Note that by (\ref{lemwkat0}) we have that $w_k(z)$ is of order $z^{1/3}$
as $z \to 0$. By (\ref{defM}) and (\ref{defM123}) this implies that
\begin{equation} \label{Mat0}
    M(z) = O(z^{-1/3}) \qquad \mbox{ as } z \to 0.
\end{equation}
The RH problem for $M$ easily gives that $\det M(z) \equiv 1$. Thus $M^{-1}(z)$
exists for $z \in \mathbb C \setminus [-z^*,z^*]$ and from (\ref{Mat0}) it
follows that $M^{-1}(z) = O(z^{-2/3})$. However, the special form of the
solution (\ref{defM})-(\ref{defM123}) shows that all cofactors of $M$ are actually
$O(z^{-1/3})$ as $z \to 0$. Thus
\begin{equation} \label{Minvat0}
    M^{-1}(z) = O(z^{-1/3}) \qquad \mbox{ as } z \to 0.
\end{equation}
This may also be understood from the fact that $M^{-1} = M^t$, since
together with $M$ it is easy to see that also $M^{-t}$ is a solution of
the RH problem (\ref{mod2})-(\ref{mod4}).

The model solution $M$ will be used to construct a {\it parametrix}
for $S$ outside of small neighborhoods of the edge
points and the origin. Namely, we consider disks of fixed radius $r$ around the
edge points and a shrinking disk $D(0, n^{-1/4})$ of radius $n^{-1/4}$ around the origin.
At the edge points and at the origin $M$ is not analytic (it is not even
bounded) and in the
disks around the edge points and the origin the parametrix
is constructed differently.

\section{Parametrix at edge points}
The construction of a parametrix $P$ at the edge points $\pm z^*$
can be done with Airy functions in a by now standard way, see
\cite{BK1,Dei,DKMVZ1,DKMVZ2}.
We omit details. We only note that $\pm z^* = \pm \frac{3\sqrt{3}}{2}c$
depends on $c$ and therefore on $a$. As $a \to 1$ we have $c \to 1$
and so $\pm z^* \to \pm \frac{3\sqrt{3}}{2}$. We construct the
Airy parametrices in fixed neighborhoods $D(\pm \frac{3\sqrt{3}}{2}, r)$
of $\pm \frac{3\sqrt{3}}{2}$ so that

\begin{itemize}
\item $P$ is analytic on $D(\pm \frac{3\sqrt{3}}{2}, r) \setminus\Sigma_S$,
\item for $z \in D(\pm \frac{3\sqrt{3}}{2}, r) \cap \Sigma_S$, we have
\begin{equation}\label{jumpP}
P_+(z)=P_-(z)j_S(z),
\end{equation}
where $j_S$ is given by (\ref{jS1})-(\ref{jS7}),
\item as $n\to\infty$,
\begin{equation}\label{asympP}
P(z)= M(z) \left(I+O\left(n^{-1} \right) \right)
\quad \textrm{ uniformly for } \left|z \pm \frac{3\sqrt{3}}{2}\right| = r.
\end{equation}
\end{itemize}

\section{Parametrix at the origin}

The main issue is the construction of a parametrix at the origin
and this is where the Pearcey integrals come in.
For $a$ sufficiently close to $1$, we want to define $Q$
in a neighborhood $D(0,r)$ of the origin such that
\begin{itemize}
\item $Q$ is analytic on $D(0,r) \setminus \Sigma_S$,
\item for $z \in D(0,r) \cap \Sigma_S$, we have
\begin{equation}\label{param1}
Q_+(z)=Q_-(z)j_S(z),
\end{equation}
where $j_S$ is given by (\ref{jS1})-(\ref{jS7}),
\item as $n\to\infty$, and with $a = 1 + O(n^{-1/2})$, we have
\begin{equation}\label{param2}
Q(z)= M(z) \left(I+O\left(n^{-1/2}\right)\right)
    \qquad \mbox{ uniformly for } |z| = n^{-1/4}.
\end{equation}
\end{itemize}
The parametrix $Q$
will be constructed with the aid of Pearcey integrals.

To motivate the construction, we note that the jump
matrices for $S$ can be factored as
\begin{equation} \label{jSfactor}
    j_S = e^{-n\Lambda_-} j_S^o e^{n \Lambda_+},
\end{equation}
where $\Lambda = \diag(\lambda_1, \lambda_2, \lambda_3)$ and
\begin{align}\label{jSo1}
j_S^o = \begin{pmatrix}
0 & 1 & 0 \\
-1& 0 & 0 \\
0 & 0 & 1
\end{pmatrix}
& \quad \textrm{ on }(0,z^*),\\ \label{jSo2}
j_S^o = \begin{pmatrix}
0 & 0 & 1 \\
0 & 1 & 0 \\
-1& 0 & 0
\end{pmatrix}
& \quad \textrm{ on } (-z^*,0),\\ \label{jSo3}
j_S^o = \begin{pmatrix}
1 & 0 & 0 \\
1 & 1 & 1 \\
0 & 0 & 1
\end{pmatrix}
& \quad \textrm{ on the upper lip of the right lens,}\\ \label{jSo4}
j_S^o = \begin{pmatrix}
1 & 0 & 0 \\
0 & 1 & 0 \\
1 & 1 & 1
\end{pmatrix}
& \quad \textrm{ on the upper lip of the left lens,}\\ \label{jSo5}
j_S^o = \begin{pmatrix}
1 & 0 & 0 \\
0 & 1 & 0 \\
1 & -1 & 1
\end{pmatrix}
& \quad \textrm{ on the lower lip of the left lens,}\\ \label{jSo6}
j_S^o = \begin{pmatrix}
1 & 0 & 0 \\
1 & 1 & -1 \\
0 & 0 & 1
\end{pmatrix}
& \quad \textrm{ on the lower lip of the right lens.}
\end{align}

We show in the next subsection that the Pearcey integrals
satisfy a RH problem with exactly the above jump matrices
except that these jumps are situated
on six rays emanating from the origin.

\subsection{The Pearcey parametrix}
\begin{figure}[h]
\includegraphics[height=5cm,width=5cm]{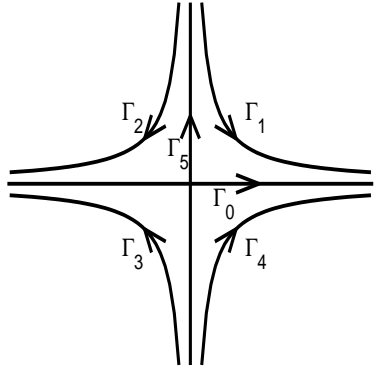}
\caption{The contours $\Gamma_j$, $j=0,1,\ldots,5$, equivalent to those in
(\ref{pi2}), that are used in the definition of the Pearcey integrals $p_j$.}
\label{fig:contoursGamma}
\end{figure}

Let $b \in \mathbb R$ be fixed. The Pearcey differential equation
$  p'''(\zeta) = \zeta p(\zeta) + b p'(\zeta)$
admits solutions of the form
\begin{equation} \label{pint}
    p_j(\zeta) = \int_{\Gamma_j} e^{-\frac{1}{4}s^4 - \frac{b}{2} s^2 +is\zeta} ds
\end{equation}
for $j=0,1,2,3,4,5$, where
\begin{equation}\label{pi2}
\begin{array}{ll}
\Gamma_0=(-\infty,\infty),\qquad &
\Gamma_1=(i\infty,0]\cup[0,\infty),\\
\Gamma_2=(i\infty,0]\cup[0,-\infty),\qquad &
\Gamma_3=(-i\infty,0]\cup[0,-\infty),\\
\Gamma_4=(-i\infty,0]\cup[0,\infty),\qquad &
\Gamma_5=(-i\infty,i\infty)
\end{array}
\end{equation}
or any other contours that are homotopic to them as for example given
in Figure \ref{fig:contoursGamma}. The formulas (\ref{pi2}) also determine the orientation of
the contours $\Gamma_j$.

Define $\Phi = \Phi(\zeta;b)$ in six sectors by
\begin{align} \label{defPhi1}
    \Phi & =
    \begin{pmatrix} - p_2 & p_1 & p_5 \\
    -p_2' & p_1' & p_5' \\
    -p_2'' & p_1'' & p_5'' \end{pmatrix}
    \qquad \textrm{ for } 0 < \arg \zeta < \pi/4 \\
    \Phi & = \label{defPhi2}
    \begin{pmatrix} p_0 & p_1 & p_4 \\
    p_0' & p_1' & p_4' \\
    p_0'' & p_1'' & p_4'' \end{pmatrix}
    \qquad \textrm{ for } \pi/4 < \arg \zeta < 3\pi/4 \\
    \Phi & = \label{defPhi3}
    \begin{pmatrix} - p_3 & -p_5 & p_4 \\
    -p_3' & -p_5' & p_4' \\
    -p_3'' & -p_5'' & p_4'' \end{pmatrix}
    \qquad \textrm{ for } 3\pi/4 < \arg \zeta < \pi \\
    \Phi & = \label{defPhi4}
    \begin{pmatrix}  p_4 & -p_5 & p_3 \\
    p_4' & -p_5' & p_3' \\
    p_4'' & -p_5'' & p_3'' \end{pmatrix}
    \qquad \textrm{ for } -\pi < \arg \zeta < -3\pi/4 \\
    \Phi & = \label{defPhi5}
    \begin{pmatrix} p_0 & p_2 & p_3 \\
    p_0' & p_2' & p_3' \\
    p_0'' & p_2'' & p_3'' \end{pmatrix}
    \qquad \textrm{ for } -3\pi/4 < \arg \zeta < -\pi/4 \\
    \Phi & = \label{defPhi6}
    \begin{pmatrix} p_1 & p_2 & p_5 \\
    p_1' & p_2' & p_5' \\
    p_1'' & p_2'' & p_5'' \end{pmatrix}
    \qquad \textrm{ for } -\pi/4 < \arg \zeta < 0
    \end{align}
Then $\Phi$ has jumps on the six rays. We choose an orientation
on these rays so that the rays in the right half-plane are oriented
from $0$ to $\infty$, and the rays in the left half-plane are
oriented from $\infty$ to $0$.

Then the integral representations (\ref{pint})-(\ref{pi2})
easily imply that
$\Phi_+ = \Phi_- j_{\Phi}$ where
\begin{align} \label{jPhi1}
j_{\Phi} = \begin{pmatrix} 1&0&0 \\ 1&1&1 \\ 0&0&1 \end{pmatrix}
    \quad \textrm{on } \arg \zeta = \pi/4, & \quad
j_{\Phi} = \begin{pmatrix} 0&1&0 \\ -1&0&0 \\ 0&0&1 \end{pmatrix}
    \quad \textrm{on } \arg \zeta = 0, \\ \label{jPhi2}
j_{\Phi} = \begin{pmatrix} 1&0&0 \\ 1&1&-1 \\ 0&0&1 \end{pmatrix}
    \quad \textrm{on } \arg \zeta = -\pi/4, & \quad
j_{\Phi} = \begin{pmatrix} 1&0&0 \\ 0&1&0 \\ 1&1&1 \end{pmatrix}
    \quad \textrm{on } \arg \zeta = 3\pi/4, \\ \label{jPhi3}
j_{\Phi} = \begin{pmatrix} 0&0&1 \\ 0&1&0 \\ -1&0&0 \end{pmatrix}
    \quad \textrm{on } \arg \zeta = -\pi, & \quad
j_{\Phi} = \begin{pmatrix} 1&0&0 \\ 0&1&0 \\ 1&-1&1 \end{pmatrix}
    \quad \textrm{on } \arg \zeta = -3\pi/4.
\end{align}
So these are indeed the jump matrices of (\ref{jSo1})-(\ref{jSo6}).

\subsection{Asymptotics of Pearcey integrals}

A classical steepest descent analysis of the integral representations
gives the following result for the  asymptotic behavior of $\Phi(\zeta;b)$ as
$\zeta \to \infty$.
As always we use the principal branches of the fractional
powers, that is, with a branch cut along the negative axis.
\begin{lem} \label{asympPearcey}
For every fixed $b \in \mathbb C$, we have as $\zeta \to \infty$,
\begin{equation} \label{Phiasymp1}
    \Phi(\zeta;b) = \sqrt{\frac{2\pi}{3}} i e^{\frac{b^2}{8}}
    \begin{pmatrix} \zeta^{-1/3} & 0 & 0 \\
        0 & 1 & 0 \\ 0 & 0 & \zeta^{1/3} \end{pmatrix}
    \begin{pmatrix} - \omega & \omega^2 & 1 \\
        -1 & 1 & 1 \\
        -\omega^2 & \omega & 1 \end{pmatrix}
        \left( I + O\left(\zeta^{-2/3} \right)\right)
    \begin{pmatrix} e^{\theta_1(\zeta;b)} & 0 & 0 \\
    0 & e^{\theta_2(\zeta;b)}& 0 \\
    0 & 0 & e^{\theta_3(\zeta;b)} \end{pmatrix}
\end{equation}
for $\Im \zeta > 0$, and
\begin{equation} \label{Phiasymp2}
    \Phi(\zeta;b) = \sqrt{\frac{2\pi}{3}} i e^{\frac{b^2}{8}}
    \begin{pmatrix} \zeta^{-1/3} & 0 & 0 \\
        0 & 1 & 0 \\ 0 & 0 & \zeta^{1/3} \end{pmatrix}
    \begin{pmatrix} \omega^2 & \omega & 1 \\
        1 & 1 & 1 \\
        \omega & \omega^2 & 1 \end{pmatrix}
        \left( I + O\left(\zeta^{-2/3} \right)\right)
    \begin{pmatrix} e^{\theta_2(\zeta;b)} & 0 & 0 \\
    0 & e^{\theta_1(\zeta;b)}& 0 \\
    0 & 0 & e^{\theta_3(\zeta;b)} \end{pmatrix}
\end{equation}
for $\Im \zeta < 0$, where $\omega = e^{2\pi i/3}$ and
\begin{equation} \label{defthetak}
    \theta_k(\zeta;b) = \frac{3}{4} \omega^{2k} \zeta^{4/3}
    + \frac{b}{2} \omega^k \zeta^{2/3}, \qquad k=1,2,3.
\end{equation}

The $O$-terms in {\rm (\ref{Phiasymp1})} and {\rm (\ref{Phiasymp2})} are
uniform for $b$ in a bounded subset of the complex plane.
\end{lem}

\begin{proof}
We give an outline of the proof; cf.\ also the calculations in \cite{Miy}.
Let $\theta(s; \zeta, b) = - \frac{1}{4} s^4 - \frac{b}{2} s^2 + i \zeta s$.
The saddle point equation for (\ref{pint}) is
\[ \frac{\partial \theta}{\partial s} = - s^3 - bs + i\zeta = 0. \]
For $b=0$ there are three solutions $s_k^o = - i \omega^k \zeta^{1/3}$, $k=1,2,3$,
and as  $\zeta \to \infty$, while $b$ remains bounded, the three
saddles $s_k = s_k(\zeta;b)$ are close to $s_k^o$, and in fact
\[ s_k(\zeta;b) = - i \omega^k \zeta^{1/3} -
    i \omega^{2k} \frac{b}{3} \zeta^{-1/3} + O(\zeta^{-5/3}) \qquad \mbox{ as } \zeta \to \infty. \]
The value at the saddles is
\[ \theta(s_k(\zeta;b); \zeta, b ) =
    \frac{3}{4} \omega^k \zeta^{4/3} + \frac{b}{2} \omega^{2k} \zeta^{2/3} + \frac{b^2}{6} + O(\zeta^{-2/3})
    \qquad \mbox{ as } \zeta \to \infty. \]

Then, if $C_k$ is the steepest descent path through $s_k$, we obtain
from classical steepest descent arguments
\[ \begin{aligned}
    \int_{C_k} e^{-\frac{1}{4}s^4 - \frac{b}{2}s^2 + i\zeta s} ds & =
    \pm \sqrt{ \frac{2\pi}{- \frac{\partial^2 \theta}{\partial s^2}(s_k(\zeta,b);\zeta,b)}} e^{\theta(s_k(\zeta;b);\zeta,b)}
        (1 + O(\zeta^{-2/3})) \\
        & = \pm
    \sqrt{ \frac{2\pi}{3}} i  \omega^{2k} \zeta^{-1/3}
    e^{\frac{3}{4} \omega^k \zeta^{4/3} + \frac{b}{2} \omega^{2k} \zeta^{2/3} + \frac{b^2}{6}}
    (1 + O(\zeta^{-2/3})).
    \end{aligned} \]
The choice of $\pm$ sign depends on the orientation of the steepest descent path.

Now take any of the six sectors that appear in the definition (\ref{defPhi1})--(\ref{defPhi6})
of $\Phi$ and take some $p_j$ that appears in the definition of $\Phi$
in that sector. The contour $\Gamma_j$ in the definition (\ref{pint}) of $p_j$
can be deformed to the steepest descent contour through one of
the saddles, or to the union of two or three such steepest descent contours. However, in the latter case,
it turns out that there is always a unique dominant saddle for $p_j$ in that particular sector. Thus
for some $k$ and some choice of $\pm$ sign, we have
\begin{equation} \label{pjasymp1}
    p_j(\zeta) =  \pm
    \sqrt{ \frac{2\pi}{3}} i  \omega^{2k} \zeta^{-1/3}
    e^{\frac{3}{4} \omega^k \zeta^{4/3} + \frac{b}{2} \omega^{2k} \zeta^{2/3} + \frac{b^2}{6}}
    (1 + O(\zeta^{-2/3})) \end{equation}
as $\zeta \to \infty$ in the chosen sector. Similarly,
\begin{align} \label{pjasymp2}
     p_j'(\zeta) & = \pm \sqrt{ \frac{2\pi}{3}} i
    e^{\frac{3}{4} \omega^k \zeta^{4/3} + \frac{b}{2} \omega^{2k} \zeta^{2/3} + \frac{b^2}{6}}
    (1 + O(\zeta^{-2/3})), \\
     p_j''(\zeta) & = \label{psasymp3}
     \pm \sqrt{ \frac{2\pi}{3}} i \omega^k \zeta^{1/3}
    e^{\frac{3}{4} \omega^k \zeta^{4/3} + \frac{b}{2} \omega^{2k} \zeta^{2/3} + \frac{b^2}{6}}
    (1 + O(\zeta^{-2/3})).
\end{align}
A further analysis reveals which value of $k$ and what sign is associated with $p_j$
in the particular sector. We will not go through this analysis here, but the result is given by
(\ref{Phiasymp1})) and (\ref{Phiasymp2}). This completes the proof of the lemma.
\end{proof}

Note that in the above lemma we only state the leading term in a full
asymptotic expansion, which is enough for the purposes of this paper.
We also stay away from situations where saddles coalesce.
For more asymptotic results on Pearcey integrals in various regimes,
see \cite{BHo,Miy,PK} and the references cited therein.

\subsection{Definition of $Q$}

We are going to define the local parametrix $Q$ in the form
\begin{equation} \label{constrdefQ}
    Q(z) = E(z) \Phi(n^{3/4} \zeta(z); n^{1/2} b(z)) e^{n \Lambda(z)} e^{-n z^2/6}, \qquad
    \Lambda = \diag(\lambda_1, \lambda_2, \lambda_3),
\end{equation}
where $E$ is an analytic prefactor, $z \mapsto \zeta(z)$ is a conformal
map from a neighborhood of $0$ in the $z$-plane to a neighborhood of
$0$ in the $\zeta$-plane, and $z \mapsto b(z)$ is analytic.

We choose $\zeta(z)$ and $b(z)$ so that the exponential
factors in the asymptotic behavior of
$\Phi(n^{3/4} \zeta(z); n^{1/2} b(z))$ are cancelled when
we multiply them by $e^{n \Lambda(z)} e^{-n z^2/6}$. We use the
functions $f_3$ and $g_3$ from Lemma \ref{lambdakat0} in
the following definition. These functions depend on $a$,
and to emphasize the $a$-dependence we write  $f_3(z;a)$ and
$g_3(z;a)$. The functions $\zeta(z)$ and $b(z)$ also depend on $a$.

\begin{defin}
For $z$ in a sufficiently small neighborhood of $0$, we define
\begin{equation} \label{defzeta}
    \zeta(z) = \zeta(z;a) = z \left[ f_3(z;a) \right]^{3/4}
\end{equation}
and
\begin{equation} \label{defb}
    b(z) = b(z;a) = \frac{g_3(z;a)}{f_3(z;a)^{1/2}}.
\end{equation}
\end{defin}
In (\ref{defzeta}) and (\ref{defb}) the branch of
the fractional powers is chosen which is real and positive
for real values of $z$ near $0$.

\begin{lem} \label{lem82}
\begin{enumerate}
\item[(a)]
There is an $r > 0$ and a $\delta > 0$  so that
for each $a \in (1-\delta, 1+\delta)$ we have that
$z \mapsto \zeta(z;a)$ is a conformal map on the disk
$D(0,r)$ and $z \mapsto b(z;a)$ is analytic on $D(0,r)$.
\item[(b)]
In addition we have
\begin{equation} \label{bzasato1}
    b(z;a) = O(a-1) + O(z^2) \quad \mbox{ as $a \to 1$ and $z \to 0$.}
    \end{equation}
\end{enumerate}
\end{lem}

\begin{proof}
Following the constructions of $f_j$ and $g_j$ for $j=1,2,3$
in Lemmas \ref{lemwkat0}, \ref{lemxikat0}, and \ref{lambdakat0}
and their proofs, we easily see that
\begin{equation} \label{corlem1}
     f_3(z;a) = f_3(z;1) + O(a-1), \quad
     g_3(z;a) = g_3(z;1) + O(a-1),
    \qquad \mbox{ as } a \to 1, \end{equation}
uniformly for $z$ in a neighborhood of $0$, and
\begin{equation} \label{corlem2}
    f_3(z;1) = 1 + O(z^2), \quad
    g_3(z;1) = O(z^2) \qquad \mbox{ as } z \to 0.
\end{equation}
Both parts of the lemma follow from (\ref{corlem1}) and (\ref{corlem2}),
and the definitions (\ref{defzeta}) and (\ref{defb}).
\end{proof}

From now on we assume that $|a-1| < \delta$, where $\delta > 0$
is as in part (a) of Lemma \ref{lem82}, so that $z \mapsto \zeta(z;a)$
is a conformal map. Near $0$ we choose the precise form of the lenses
so that the lips of the lenses are mapped by
$z \mapsto \zeta(z;a)$ to the rays $\arg \zeta = \pm \pi/4$
and $\arg \zeta = \pm 3\pi/4$.
Then from the fact that the jump matrices (\ref{jPhi1})--(\ref{jPhi3})
of $\Phi$ agree with those in (\ref{jSo1})--(\ref{jSo6}),
it follows that the jump condition (\ref{param1}) for  $Q$ is satisfied.
This holds for any choice of analytic prefactor $E$
that is used in (\ref{constrdefQ}) to define $Q$.
We are going to define $E$ so that the matching condition (\ref{param2})
is satisfied as well.

\subsection{Matching condition}

To obtain the matching condition (\ref{param2}) we first note
that the definitions (\ref{defzeta}) and (\ref{defb})
give us (we drop the $a$-dependence in the notation)
\[ \zeta(z)^{4/3} = z^{4/3} f_3(z), \qquad b(z) \zeta(z)^{2/3} = g_3(z). \]
Hence by (\ref{lambdakf3g3}) and (\ref{defthetak}) we have
for $\Im z > 0$ with $|z| < r$,
\begin{equation} \label{thetaklambdak1}
    \theta_k(\zeta(z);b(z)) + \lambda_k(z) - z^2/6 = 0,
    \qquad k =1,2,3,
\end{equation}
while for $\Im z < 0$ with $|z| < r$,
\begin{equation} \label{thetaklambdak2}
\begin{aligned}
    \theta_2(\zeta(z);b(z)) + \lambda_1(z) - z^2/6 &= \lambda_{1-}(0) = \pi i,  \\
    \theta_1(\zeta(z);b(z)) + \lambda_2(z) - z^2/6 &= \lambda_{2-}(0) = -\pi i, \\
    \theta_3(\zeta(z);b(z)) + \lambda_3(z) - z^2/6 &= \lambda_{3-}(0) = 0.
    \end{aligned} \end{equation}

Assume $a = 1 + O(n^{-1/2})$. Then it follows from
(\ref{bzasato1}) that
\begin{equation} \label{bsmall}
    n^{1/2} |b(z;a)| \leq C     \qquad \mbox{ for } |z| \leq 2 n^{-1/4}
\end{equation}
for every $n$ large enough, with a value $C$ that is independent of $n$.
As a consequence we can use the expansions (\ref{Phiasymp1}), (\ref{Phiasymp2})
as $n \to \infty$, because of Lemma \ref{asympPearcey}.
We find from (\ref{Phiasymp1}) and (\ref{Phiasymp2}) and the relations
(\ref{thetaklambdak1}) and (\ref{thetaklambdak2}) between
$\theta_k$ and $\lambda_k$ that
the exponential factors in the asymptotic behavior (as $n \to \infty$) of
\[ \Phi(n^{3/4} \zeta(z); n^{1/2} b(z)) e^{n\Lambda} e^{-n z^2/6} \]
cancel if we take $z$ so that $0.9 n^{-1/4} \leq |z| \leq 1.1 n^{-1/4}$.
So we have proved the following.
\begin{lem} Let $a = 1 + O(n^{-1/2})$.
Then we have as $n \to \infty$, uniformly for $z$ so that
$0.9 n^{-1/4} \leq |z| \leq 1.1 n^{-1/4}$,
that
\begin{equation} \label{Qasymp1} \begin{aligned}
    Q(z) & = E(z) \Phi(n^{3/4} \zeta(z); n^{1/2} b(z))
        e^{n \Lambda(z)} e^{-n z^2/6} \\
    &= \sqrt{\frac{2\pi}{3}} i e^{n b(z)^2/8}
        E(z) \begin{pmatrix} n^{-1/4}&0&0\\0&1&1\\0&0&n^{1/4} \end{pmatrix}
         K(\zeta(z)) ( I + O(n^{-1/3}))
         \end{aligned} \end{equation}
 where
 \begin{equation} \label{defK} K(\zeta) = \left\{
    \begin{array}{ll}
    \begin{pmatrix} \zeta^{-1/3}&0&0\\0&1&0\\0&0&\zeta^{1/3} \end{pmatrix}
    \begin{pmatrix} -\omega & \omega^2& 1\\
    -1 & 1 & 1 \\ -\omega^2 & \omega & 1\end{pmatrix} &
    \qquad \mbox{ for } \Im \zeta > 0, \\[20pt]
    \begin{pmatrix} \zeta^{-1/3}&0&0\\0&1&0\\0&0&\zeta^{1/3} \end{pmatrix}
    \begin{pmatrix} \omega^2 & \omega& 1\\
    1 & 1 & 1 \\ \omega & \omega^2 & 1\end{pmatrix} &
    \qquad \mbox{ for } \Im \zeta < 0.
    \end{array} \right.
\end{equation}
\end{lem}

\begin{proof}
This follows from the asymptotic behavior (\ref{Phiasymp1}) and (\ref{Phiasymp2}),
since we have shown in the above that the exponential factors in (\ref{Phiasymp1})
and (\ref{Phiasymp2}) are cancelled when we multiply them by $e^{n \Lambda(z)} e^{-n z^2/6}$.

As for the $O$-term, we note that $n^{3/4}\zeta(z) = O(n^{1/2})$ if $|z| = cn^{-1/4}$
with $0.9 \leq c \leq 1.1$,
so that the $O(\zeta^{-2/3})$ term in (\ref{Phiasymp1})-(\ref{Phiasymp2})
leads to the $O(n^{-1/3})$ term in (\ref{Qasymp1}).
\end{proof}

In order to achieve the matching (\ref{param2}) of $Q(z)$ with $M(z)$ we
now {\em define} the prefactor $E$ by
\begin{equation} \label{defE}
    E(z) = -\sqrt{\frac{3}{2\pi}} i e^{-n b(z)^2/8}
    M(z) K(\zeta(z))^{-1}
    \begin{pmatrix} n^{1/4}&0&0\\0&1&0\\0&0&n^{-1/4}\end{pmatrix}.
    \end{equation}
Then the matching condition (\ref{param2}) follows from
(\ref{Qasymp1}) and (\ref{defE}).

It only remains to check that $E$ is analytic in a full
neighborhood of the origin.
This follows since $M$ and $K$ satisfy the
same jump relations on the real line. Indeed we have from the
expressions (\ref{defK}) for $K$, for real $\zeta$ with $\zeta > 0$,
\[ K_-(\zeta)^{-1} K_+(\zeta)
    =  \begin{pmatrix} \omega^2 & \omega& 1\\
    1 & 1 & 1 \\ \omega & \omega^2 & 1\end{pmatrix}^{-1}
    \begin{pmatrix} -\omega & \omega^2& 1\\
    -1 & 1 & 1 \\ -\omega^2 & \omega & 1\end{pmatrix}
    = \begin{pmatrix} 0&1&0\\-1&0&0\\0&0&1\end{pmatrix}
    \]
while for real $\zeta < 0$ we have to take into account
that $\zeta^{1/3}$ and $\zeta^{-1/3}$ have different $\pm$-boundary values,
so that for $\zeta < 0$,
\[ \begin{aligned}
    K_-(\zeta)^{-1} K_+(\zeta)
    &
    =  \begin{pmatrix} \omega^2 & \omega& 1\\
    1 & 1 & 1 \\ \omega & \omega^2 & 1\end{pmatrix}^{-1}
    \begin{pmatrix} \zeta_-^{1/3}\zeta_+^{-1/3}&0&0\\0&1&0\\
    0&0&\zeta_-^{-1/3}\zeta_+^{1/3}\end{pmatrix}
    \begin{pmatrix} -\omega & \omega^2& 1\\
    -1 & 1 & 1 \\ -\omega^2 & \omega & 1\end{pmatrix} \\
    &=\begin{pmatrix} \omega^2 & \omega& 1\\
    1 & 1 & 1 \\ \omega & \omega^2 & 1\end{pmatrix}^{-1}
    \begin{pmatrix} \omega^2 &0&0\\0&1&0\\
    0&0&\omega \end{pmatrix}
    \begin{pmatrix} -\omega & \omega^2& 1\\
    -1 & 1 & 1 \\ -\omega^2 & \omega & 1\end{pmatrix}
    = \begin{pmatrix} 0&0&1\\0&1&0\\-1&0&0\end{pmatrix}.
    \end{aligned}
    \]
These are indeed equal to the jumps satisfied by $M$; see (\ref{mod2}) and (\ref{mod3}).
Since $\zeta(z)$ is a conformal map on $D(0,r)$ that
is real and positive for $z \in (0,r)$,
and real and negative for $z \in (-r,0)$,
we find that $M(z) K(\zeta(z))^{-1}$ is analytic
across both $(0,r)$ and $(-r,0)$. Thus $E(z)$
is analytic in $D(0,r) \setminus \{0\}$.
The isolated singularity at $0$ is removable, since
the entries in $M(z)$ and $K(\zeta(z))^{-1}$ have at most
$z^{-1/3}$-type singularity at the origin, and they
cannot combine to form a pole.
The conclusion is that $E$ is analytic.

This completes the construction of the local parametrix
$Q$ at the origin.

\section{Final transformation}

\begin{figure}[h]
\includegraphics[height=1.3cm,width=12cm]{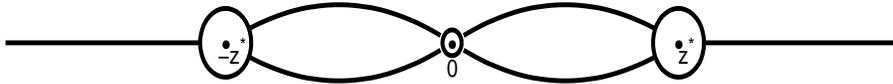}
\caption{The contour $\Sigma_R$. The matrix-valued function $R$ is analytic
on $\mathbb C \setminus \Sigma_R$. The disk around $0$ has radius $n^{-1/4}$
and is shrinking as $n \to \infty$. The disks are oriented counterclockwise and the
remaining parts of $\Sigma_R$ are oriented from left to right.}
\label{fig7}
\end{figure}

We now fix $b \in \mathbb R$ and let $a = 1 + \frac{b}{2\sqrt{n}}$.
Now we define
\begin{equation} \label{defR}
    R(z) =
    \left\{ \begin{array}{ll}
        S(z) M(z)^{-1}, & \textrm{ for $z \in \mathbb C \setminus \Sigma_S$ outside the disks
        $D(0,n^{-1/4})$ and $D(\pm \frac{3\sqrt{3}}{2},r)$}, \\[10pt]
        S(z) P(z)^{-1}, & \textrm{ for } z \in D(\pm \frac{3\sqrt{3}}{2},r) \setminus \Sigma_S , \\[10pt]
        S(z) Q(z)^{-1}, & \textrm{ for } z \in D(0, n^{-1/4}) \setminus \Sigma_S.
        \end{array} \right.
        \end{equation}
Then $R$ is analytic inside the disks and also across the real interval
between the disks. Thus $R$ is analytic outside the  contour $\Sigma_R$
shown in Figure \ref{fig7}.

\begin{lem} \label{lemma_jR} We have $R_+ = R_-j_R$ where
\begin{align} \label{jR1}
    j_R(z) &= I + O(n^{-1}) \qquad \mbox{uniformly for } \left|z \mp \frac{3\sqrt{3}}{2}\right| = r,  \\
    \label{jR2}
   j_R(z)  &= I + O(n^{-1/6}) \quad \mbox{ uniformly for } |z| = n^{-1/4},
   \end{align}
and there exists $c > 0$ so that
\begin{align}
   \label{jR3}
    j_R(z) & = I + O\left(\frac{e^{-cn^{2/3}}}{1+|z|^2}\right) \quad \mbox{ uniformly for $z$ on
    the remaining parts of $\Sigma_R$.}
    \end{align}
\end{lem}
\begin{proof}
The behavior (\ref{jR1}) of the jump matrix  on the circles around
the endpoints $\pm \frac{3\sqrt{3}}{2}$ is a result of the
construction of the Airy parametrix. It follows as in
\cite{DKMVZ1,DKMVZ2}.

The jump matrix  for $|z| = n^{-1/4}$ is by (\ref{defR})
and (\ref{param2})
(we use positive  orientation)
\[ j_R = M Q{-1} = M(I + O(n^{-1/3})) M^{-1} =
    M + M O(n^{-1/3}) M^{-1}(z). \]
Since $M(z) = O(z^{-1/3})$ and $M^{-1}(z) = O(z^{-1/3})$ as $z \to 0$
by (\ref{Mat0}) and (\ref{Minvat0}), we obtain (\ref{jR2}).

The jump matrix $j_R(z)$ on the remaining part of $\Sigma_R$ is
$I + O(e^{-cn})$ if $z \in \Sigma_R$ stays at a fixed distance
of $0$ and $\pm \frac{3\sqrt{3}}{2}$. But now the disk around
$0$ is shrinking as $n$ increases, and so we have to be more careful here.
We note that the jump matrix is
\[ j_R(z) = M(z) j_S(z) M^{-1}(z) \]
and we want to know its behavior as $n \to \infty$
for $z$ on the lips of the lenses near $0$ and $|z| \geq n^{-1/4}$.

The jump matrices $j_S$ in (\ref{jS4})--(\ref{jS7}) contain off-diagonal
entries $\pm e^{n(\lambda_k-\lambda_j)}$.
For $a=1$ these entries are decaying on the contours and so we have
for some positive constant $c_1 > 0$.
\[ \Re((\lambda_j-\lambda_k)(z;1)) \geq c_1 |z|^{4/3} \]
for $z$ on the lips of the lenses near $0$.
Since $\lambda_j(z;a) = \lambda_j(z;1) + z^{2/3} O(a-1)$
as $a \to 1$, we then get that
\[ \Re((\lambda_j-\lambda_k)(z;a)) \geq c_1 z^{-4/3} -
    c_2 |z|^{2/3}|a-1|. \]
Then if $a-1 = (b/2) n^{-1/2}$ and $|z| \geq n^{-1/4}$
we easily get that
\[ \Re(\lambda_j-\lambda_k)(z;a)) \geq c_3 n^{-1/3} \]
for some positive constant $c_3 >0$. Then it follows from
(\ref{jS4})--(\ref{jS7}) that
\[ j_S(z) = I + O(e^{-c_3 n^{2/3}}). \]
This leads to (\ref{jR3}) since $M(z) = O(z^{-1/3})$ and $M^{-1}(z) = O(z^{-1/3})$
as $z \to 0$, see (\ref{Mat0}) and (\ref{Minvat0}).
\end{proof}

To summarize, we find that $R$ solves the following RH problem:
\begin{itemize}
\item $R$ is analytic on $\C\setminus\Sigma_R$,
\item for $z \in \Sigma_R$, we have $R_+=R_-j_R$,
where $j_R$ satisfies (\ref{jR1})-(\ref{jR3}),
\item as $z\to\infty$, we have $R(z)=I+O(1/z)$.
\end{itemize}

The RH problem for $R$ is posed on a contour that is varying with $n$.
This is a slight complication. However we still can guarantee
the following behavior of $R$ as $n \to \infty$.

\begin{prop}\label{Rz}
As $n \to \infty$ we have that
\begin{equation} \label{asympR1}
    R(z) = I + O\left(\frac{n^{-1/6}}{1+|z|}\right)
\end{equation}
uniformly for $z \in \mathbb C \setminus \Sigma_R$.
\end{prop}

Since the proof of Proposition \ref{Rz} is somewhat technical due to the fact
that the contours are varying with $n$, we give it in Appendix \ref{App_A}.

\section{Proof of Theorem \ref{maintheo}}
Now we are ready for the proof of Theorem \ref{maintheo}.
We fix $b \in \mathbb R$ and take
\[ a = 1 + \frac{b}{2\sqrt{n}}. \]

\subsection{The effect of the transformations $Y \mapsto T \mapsto S \mapsto R$}
We are going to follow the effect of the transformations on
the correlation kernel $K_n(x,y;a)$ for real values of $x$ and $y$
close to $0$. We start from
(\ref{corrkernel}) which gives $K_n(x,y;a)$ in terms of the
solution of the RH problem for $Y$. The transformation (\ref{defT}) then implies that
\begin{equation} \label{corrkernel2}
    K_n(x,y;a) = \frac{e^{\frac{1}{4}n(x^2-y^2)}}{2\pi i(x-y)}
        \begin{pmatrix} 0 & e^{n\lambda_{2+}(y)} & e^{n \lambda_{3+}(y)} \end{pmatrix}
            T_+^{-1}(y) T_+(x) \begin{pmatrix} e^{-n \lambda_{1+}(x)} \\ 0 \\ 0 \end{pmatrix}.
            \end{equation}
According to the transformation $T \mapsto S$ given in (\ref{defS1})--(\ref{defS4})
we now have to distinguish between $x$ and $y$ being positive or negative.
We will do the calculations explicitly for $x > 0$ and $y<0$.
The other cases are treated in the same way.

So we assume that $x > 0$ and $y < 0$, and both of them are close to $0$.
The formulas (\ref{defS1}) and (\ref{defS3}) applied to (\ref{corrkernel2})
then give
\begin{equation} \label{corrkernel3}
    K_n(x,y;a) = \frac{e^{\frac{1}{4}n(x^2-y^2)}}{2\pi i(x-y)}
        \begin{pmatrix} - e^{n \lambda_{1+}(y)} & 0 & e^{n \lambda_{3+}(y)} \end{pmatrix}
            S_+^{-1}(y) S_+(x)
    \begin{pmatrix} e^{-n \lambda_{1+}(x)} \\ e^{-n \lambda_{2+}(x)} \\ 0 \end{pmatrix}.
\end{equation}

Next we note that for $z$ close to $0$, inside the disk or radius
$n^{-1/4}$, we have by (\ref{defR}),
\[ S(z) = R(z) Q(z) = R(z) \widetilde{Q}(z) e^{n \Lambda(z)} e^{-n z^2/6} \]
where
\begin{equation} \label{deftildeQ}
    \widetilde{Q}(z) = Q(z) e^{-n \Lambda(z)} e^{n z^2/6} =
    E(z) \Phi(n^{3/4} \zeta(z;a); n^{1/2} b(z;a));
\end{equation}
see (\ref{Qasymp1}). Thus if $0 < x < n^{-1/4}$ and $-n^{-1/4} < y < 0$,
we have
\[ S_+(x) \begin{pmatrix} e^{-n \lambda_{1+}(x)} \\ e^{-n \lambda_{2+}(x)} \\ 0 \end{pmatrix}
     = R(x) \widetilde{Q}(x) \begin{pmatrix} 1 \\ 1 \\ 0 \end{pmatrix} e^{-n x^2/6} \]
and
\[    \begin{pmatrix} - e^{n \lambda_{1+}(y)} & 0 & e^{n \lambda_{3+}(y)} \end{pmatrix}
            S_+^{-1}(y) =
          e^{n y^2/6}  \begin{pmatrix} -1 & 0& 1 \end{pmatrix}
            \widetilde{Q}^{-1}(y) R^{-1}(y). \]
Inserting these two relations into (\ref{corrkernel3}) we find that
\begin{equation} \label{corrkernel4}
    K_n(x,y;a) = \frac{e^{\frac{1}{12}n(x^2-y^2)}}{2\pi i(x-y)}
        \begin{pmatrix} -1 & 0 & 1 \end{pmatrix}
            \widetilde{Q}^{-1}_+(y) R^{-1}(y) R(x) \widetilde{Q}_+(x)
        \begin{pmatrix} 1 \\ 1 \\ 0 \end{pmatrix}.
\end{equation}

To obtain the scaling limit (\ref{limitkernel}) of $K_n$ we
need the following lemma.
\begin{lem} \label{limits}
Let $a_n = 1 + (b/2) n^{-1/2}$.
\begin{enumerate}
\item[(a)] Let $x_n = x n^{-3/4}$ where $x \in \mathbb R$ is fixed.
Then
\begin{equation} \label{limitzeta}
    \lim_{n\to\infty}  n^{3/4} \zeta(x_n; a_n) = x
    \end{equation}
and
\begin{equation} \label{limitb}
    \lim_{n \to\infty} n^{1/2} b(x_n; a_n) = b.
\end{equation}
\item[(b)]
Let also $y_n = y n^{-3/4}$ where $y \in \mathbb R$ is fixed.
Then
\begin{equation} \label{limitRE}
   \lim_{n \to \infty} E^{-1}(y_n) R^{-1}(y_n) R(x_n) E(x_n) = I
\end{equation}
\end{enumerate}
\end{lem}
\begin{proof}
(a) Since $\zeta(z;a) = z \left[f_3(z;a)\right]^{3/4}$ by (\ref{defzeta})
and $f_3(z;a) \to 1$ if $z \to 0$ and $a \to 1$
by (\ref{corlem1}) and (\ref{corlem2}), we get that the limit
(\ref{limitzeta}) immediately follows.

For (\ref{limitb}) we need to go back to the definitions
in Lemmas \ref{lemwkat0}, \ref{lemxikat0}, and \ref{lambdakat0}
of $f_1$ and $g_j$, for $j=1,2,3$.
From \ref{lemwkat0} and its proof it follows that
$f_1(z;a) = f_1(0;a) + O(z^2)$ and $g_1(z;a) = g_1(0) + O(z^2)$
as $z \to 0$, and the $O$-terms are uniform with respect
to $a$ in a neighborhood of $1$.
Then by (\ref{defg2}) we have
\begin{equation} \label{g2near0}
    g_2(z;a) = g_2(0;a) + O(z^2) \qquad \mbox{ as } z \to 0
\end{equation}
uniformly for $a$ in a neighborhood of $1$.
Since we have; cf.\ Lemmas \ref{lemxikat0} and (\ref{lambdakat0},
\[ \frac{1}{2} z^{2/3} g_3(z;a) = \int_0^z s^{-1/3} g_2(s;a) ds \]
we get from (\ref{g2near0}) that
\begin{equation} \label{g3near0}
    g_3(z;a) = g_3(0;a) + O(z^2) \qquad \mbox{ as } z \to 0
    \end{equation}
again uniformly for $a$ in a neighborhood of $1$.
By (\ref{f3g3at0}) we have $g_3(0;a) = 3c^{-2/3}(c^2-1)$
where $c = (a+ \sqrt{a^2+8})/4 = (a-1)/3 + O((a-1)^2)$
as $a \to 1$. Thus $g_3(0;a) = 2(a-1) + O((a-1)^2)$ as
$a \to 1$, and it follows from (\ref{g3near0}) and the
definitions of $a_n$ and $x_n$ that
\[ n^{1/2} g_3(x_n; a_n) =
    n^{1/2} g_3(0;a_n) + O(n^{-1}) =
    2(a_n-1) + O(n^{-1/2}) = b + O(n^{-1/2})
    \qquad \mbox{ as } n \to\infty. \]
Then (\ref{limitb}) follows because of the definition (\ref{defb})
and the fact that $f_3(x_n;a_n) \to 1$ as $n \to \infty$.

\medskip

(b) Since $M(z) K(\zeta(z))^{-1}$ is analytic in a
neighborhood of the origin and $x_n - y_n = O(n^{-3/4})$, we have
\[ K(\zeta(y_n)) M(y_n)^{-1} M(x_n) K(\zeta(x_n))^{-1} = I + O(n^{-3/4}) \]
as $n \to \infty$. Hence by (\ref{defE})
\begin{equation} \label{limitE2}
    \begin{aligned}
    E^{-1}(y_n) E(x_n) &= e^{n(b(y_n;a_n)^2 - b(x_n;a_n)^2)/8}
        \begin{pmatrix} n^{-1/4} & 0 & 0 \\ 0 & 1 & 0 \\ 0 & 0 & n^{1/4} \end{pmatrix}
        (I + O(n^{-3/4}))
        \begin{pmatrix} n^{1/4} & 0 & 0 \\ 0 & 1 & 0 \\ 0 & 0 & n^{-1/4} \end{pmatrix} \\
        & = e^{n(b(y_n;a_n)^2 - b(x_n;a_n)^2)/8}
        (I + O(n^{-1/4})).
        \end{aligned} \end{equation}
Note that both $nb(y_n;a_n)^2$ and $nb(x_n;a_n)^2$ tend
to $b^2$ as $n \to \infty$ because of (\ref{limitb}). Thus
we also get from (\ref{defE}) that
\begin{equation} \label{asympE}
    E(x_n) = O(n^{1/4}), \qquad \mbox{and} \qquad E^{-1}(y_n) = O(n^{1/4}).
\end{equation}

Next, we get from (\ref{asympR1}) and Cauchy's theorem that
for $z = O(n^{-3/4})$ we have
\[ \frac{d}{dz} R(z) =
    \frac{1}{2\pi i} \int_{|s| = n^{-1/4}} \frac{R(s)}{s-z} ds
        = O(n^{-1/6})
        \qquad \mbox{ as } n \to \infty. \]
Then by the mean-value theorem,
\[ R(x_n) - R(y_n) = O((x_n - y_n) n^{-1/6}) = O(n^{-11/12}) \]
so that
\begin{equation} \label{asympR2}
    R^{-1}(y_n) R(x_n) = I + R^{-1}(y_n)(R(x_n) - R(y_n))
    = I + O(n^{-11/12}).
\end{equation}

Combining (\ref{limitb}), (\ref{limitE2}), (\ref{asympE}) and (\ref{asympR2})
we obtain (\ref{limitRE}).
\end{proof}

Now we can compute the double scaling limit of $K_n$.
Indeed, it follows from (\ref{deftildeQ}), (\ref{corrkernel4}), and
Lemma \ref{limits} that
\begin{equation} \label{Knlimit}
    \lim_{n \to \infty}
    \frac{1}{n^{3/4}} K_n\left(\frac{x}{n^{3/4}}, \frac{y}{n^{3/4}};
    1 + \frac{b}{2\sqrt{n}} \right)
        = K^{cusp}(x,y;b)
     \end{equation}
where
\begin{equation} \label{Kcusp1}
     K^{cusp}(x,y;b) = \frac{1}{2\pi i(x-y)}
        \begin{pmatrix} -1 & 0 & 1 \end{pmatrix}
            \Phi^{-1}_+(y;b) \Phi_+(x;b)
        \begin{pmatrix} 1 \\ 1 \\ 0 \end{pmatrix}
        \quad \mbox{if $x > 0$ and $y < 0$.}
\end{equation}
Similar calculations show that the limit (\ref{Knlimit})
exists for all $x$ and $y$, and
\begin{equation} \label{Kcusp2}
     K^{cusp}(x,y;b) = \frac{1}{2\pi i(x-y)}
        \begin{pmatrix} -1 & 1 & 0 \end{pmatrix}
            \Phi^{-1}_+(y;b) \Phi_+(x;b)
        \begin{pmatrix} 1 \\ 1 \\ 0 \end{pmatrix}
       \quad \mbox{if $x > 0$ and $y > 0$,}
\end{equation}
\begin{equation} \label{Kcusp3}
     K^{cusp}(x,y;b) = \frac{1}{2\pi i(x-y)}
        \begin{pmatrix} -1 & 1 & 0 \end{pmatrix}
            \Phi^{-1}_+(y;b) \Phi_+(x;b)
        \begin{pmatrix} 1 \\ 0 \\ 1 \end{pmatrix}
          \qquad \mbox{if $x < 0$ and $y > 0$,}
\end{equation}
\begin{equation} \label{Kcusp4}
     K^{cusp}(x,y;b) = \frac{1}{2\pi i(x-y)}
        \begin{pmatrix} -1 & 0 & 1 \end{pmatrix}
            \Phi^{-1}_+(y;b) \Phi_+(x;b)
        \begin{pmatrix} 1 \\ 0 \\ 1 \end{pmatrix}
          \qquad \mbox{if $x < 0$ and $y < 0$.}
\end{equation}

\subsection{Different formula for $K^{cusp}$}

To complete the proof of Theorem \ref{maintheo} we show that
the formulas (\ref{Kcusp1})--(\ref{Kcusp4})
for $K^{cusp}$ can be rewritten in the form
(\ref{Pearceykernel}) given in the theorem. This involves the
Pearcey integrals $p(x)$ and $q(y)$ of (\ref{Pearceyintegrals}).

Define
\begin{equation} \label{deftildePhi}
    \widetilde{\Phi} =
    \begin{pmatrix} p_0 & p_1 & p_4 \\
        p_0' & p_1'& p_4' \\
        p_0'' & p_1'' & p_4''
        \end{pmatrix}.
\end{equation}
Then by (\ref{defPhi2}) we have that $\widetilde{\Phi}(\zeta)$
agrees with $\Phi(\zeta)$ in the sector $\pi/4 < \arg \zeta < 3 \pi/4$,
but (\ref{deftildePhi}) defines $\widetilde{\Phi}$ in the
full complex $\zeta$-plane, and in particular on the real axis.

Using the jump relation $\Phi_+ = \Phi_- j_{\Phi}$
for $\arg \zeta = \pi/4$ and $\arg \zeta = 3\pi/4$,
see (\ref{jPhi1}) and (\ref{jPhi2}), we find that
\[ \Phi_+(x;b) = \widetilde{\Phi}(x;b)
    \begin{pmatrix} 1 & 0 & 0 \\
    -1 & 1 & -1 \\ 0 & 0 & 1 \end{pmatrix}
    \qquad \mbox{ if } x > 0 \]
and
\[ \Phi_+(x;b) = \widetilde{\Phi}(x;b)
    \begin{pmatrix} 1 & 0 & 0 \\
    0 & 1 & 0 \\  -1 & -1 & 1 \end{pmatrix}
    \qquad \mbox{ if } x < 0. \]
Inserting this into (\ref{Kcusp1})-(\ref{Kcusp4}) we find that
all four cases lead to
\begin{equation} \label{Kcusp5}
     K^{cusp}(x,y;b) = \frac{1}{2\pi i(x-y)}
        \begin{pmatrix} 0 & 1 & 1 \end{pmatrix}
            \widetilde{\Phi}^{-1}(y;b)
            \widetilde{\Phi}(x;b)
        \begin{pmatrix} 1 \\ 0 \\ 0 \end{pmatrix}
\end{equation}
which is the same expression for all $x, y \in \mathbb R$.

Our next task is compute $\widetilde{\Phi}^{-1}$. The inverse of
$\widetilde{\Phi}$ is built out of solutions of
\begin{equation} \label{dvq}
    q'''(z) = - z q(z) + b q'(z).
\end{equation}
It is easy to see that for any solution $q$ of (\ref{dvq})
and any solution $p$ of the Pearcey equation
\begin{equation} \label{dvp}
    p'''(z) = zp(z) + bp'(z)
\end{equation} we
have $\left(pq'' - p'q' + p''q - bpq \right)' = 0$
so that
\[ [p,q] := pq'' - p'q' + p''q - bpq = const. \]
It follows that each row of $\widetilde{\Phi}^{-1}$ has the form
$ \begin{pmatrix} q''-bq & -q' & q \end{pmatrix}$
for some particular solution of (\ref{dvq}).
More precisely, since $\widetilde{\Phi}$ is given by (\ref{deftildePhi}),
we have
\begin{equation} \label{deftildePhiinverse}
    \widetilde{\Phi}^{-1} =
    \begin{pmatrix} q_1''-bq_1 & -q_1' & q_1 \\
    q_2''-bq_2 & -q_2' & q_2 \\
    q_3''-bq_3 & -q_3' & q_3
    \end{pmatrix} \end{equation}
where
\begin{equation} \label{pqrelations}
\begin{aligned} {}
    [p_0,q_1] = 1, \quad & [p_1,q_1] = 0, & [p_4,q_1]=0, \\
    [p_0,q_2] = 0, \quad & [p_1,q_2] = 1, & [p_4,q_2]=0, \\
    [p_0,q_3] = 0, \quad & [p_1,q_3] = 0, & [p_4,q_3]=0,
    \end{aligned}
    \end{equation}
Then if $q_0 = q_2 + q_3$ we have
\begin{equation}
    [p_0,q_0] = 0, \quad [p_1, q_0] = 1, \quad [p_4,q_0] = 1,
\end{equation}
and from (\ref{deftildePhi}), (\ref{Kcusp5}), and (\ref{deftildePhiinverse})
it follows that
\begin{equation} \label{Kcusp6}
     K^{cusp}(x,y;b) =
        \frac{p_0(x)q_0(y) - p_0'(x)q_0'(y) + p_0''(x)q_0(y) - bp_0(x) q_0(y)}
        {2\pi i(x-y)}.
\end{equation}

Recall that (\ref{dvp}) has solutions with integral representations
\begin{equation} \label{pint2}
    p(z) = \int_{\Gamma} e^{-\frac{1}{4} s^4 - \frac{b}{2} s^2 + isz} ds
\end{equation}
where $\Gamma$ is a contour in the complex plane that starts and
ends at infinity at one of the angles $0$, $\pm \pi/2$, or $\pi$.
Similarly, there are solutions of (\ref{dvq}) with integral representation
\begin{equation} \label{qint}
    q(z) = \frac{1}{2\pi i} \int_{\Sigma} e^{\frac{1}{4} t^4 + \frac{b}{2} t^2 + itz} dt
\end{equation}
where $\Sigma$ is a contour in the complex plane that starts and
ends at infinity at one of the angles $\pm \pi/4$ or $\pm 3\pi/4$.

\begin{lem} \label{GammaSigma}
Let $p$ and $q$ be given by {\rm(\ref{pint2})} and {\rm(\ref{qint})}
such that $\Gamma \cap \Sigma = \emptyset$. Then
\[ [p,q] = 0. \]

If $\Gamma \cap \Sigma  = \{ z_0\}$ and $\Gamma$ and $\Sigma$
intersect transversally at $z_0$, and if the contours
are oriented so that $\Gamma$ meets $\Sigma$ in $z_0$ on the $-$-side
of $\Sigma$, then
\[ [p,q] = 1 \]
\end{lem}
\begin{proof}
We write $[p,q] = pq'' - p'q' + p''q - bpq$ as a double integral,
and for convenience we take $z = 0$. So from (\ref{pint2}) and (\ref{qint}),
\[ \begin{aligned}
   {}[p,q] & =
    \frac{1}{2\pi i} \int_{\Sigma} \int_{\Gamma}
    (-t^2 - st - s^2 - b)
    e^{\frac{1}{4} t^4 + \frac{b}{2} t^2 -\frac{1}{4} s^4 - \frac{b}{2} s^2} ds dt \\
    & =
    \frac{1}{2\pi i} \int_{\Sigma} \int_{\Gamma}
    \frac{t^3 + bt - s^3 - bs}{s-t}
    e^{\frac{1}{4} t^4 + \frac{b}{2} t^2 -\frac{1}{4} s^4 - \frac{b}{2} s^2} ds dt
\end{aligned} \]
If $\Gamma \cap \Sigma = \emptyset$ then we can write this as
\[ \begin{aligned}
   {}[p,q]     = &
    \frac{1}{2\pi i} \int_{\Sigma} \int_{\Gamma}
        \frac{1}{s-t} e^{\frac{1}{4} t^4 + \frac{b}{2} t^2} \frac{\partial}{\partial s}
        \left[ e^{ -\frac{1}{4} s^4 - \frac{b}{2} s^2}\right] ds dt \\
     & + \frac{1}{2\pi i} \int_{\Gamma} \int_{\Sigma}
        \frac{1}{s-t} e^{-\frac{1}{4} s^4 - \frac{b}{2} s^2} \frac{\partial}{\partial t}
        \left[ e^{ \frac{1}{4} t^4 + \frac{b}{2} t^2}\right] dt ds
     \end{aligned}
    \]
and we can apply integration by parts to both inner integrals. The integrated
terms vanish because of the choice of contours and the result is
\[ \begin{aligned}
   {}[p,q] & =  -
    \frac{1}{2\pi i} \int_{\Sigma} \int_{\Gamma}
        e^{\frac{1}{4} t^4 + \frac{b}{2} t^2 -\frac{1}{4} s^4 - \frac{b}{2} s^2}
            \frac{\partial}{\partial s} \left[\frac{1}{s-t} \right]
        ds dt \\
     & \quad - \frac{1}{2\pi i} \int_{\Gamma} \int_{\Sigma}
        e^{\frac{1}{4} t^4 + \frac{b}{2} t^2 -\frac{1}{4} s^4 - \frac{b}{2} s^2}
            \frac{\partial}{\partial t} \left[\frac{1}{s-t} \right] dt ds
       = 0.
     \end{aligned}
    \]

If $\Gamma \cap \Sigma \neq \emptyset$ then we cannot make the splitting of
integrals as above, and we have to proceed differently. If $\Gamma$ and $\Sigma$
intersect at $z_0$  as in the statement of the second part of the lemma, then we can deform
contours so that $\Gamma$ and $\Sigma$ intersect in $0$, and that for some $\delta >0$,
$\Sigma$ contains the real interval $[-\delta, \delta]$ oriented
from left to right, and $\Gamma$ contains  the vertical interval $[-i \delta, i \delta]$
oriented from bottom to top. Let $\varepsilon \in (0, \delta)$ and write
$\Sigma_{\varepsilon} = \Sigma \setminus (-\varepsilon, \varepsilon)$.
Then it follows as above that
\[ \begin{aligned}
   {}[p,q] & = \lim_{\varepsilon \to 0} \frac{1}{2\pi i}
    \int_{\Sigma_{\varepsilon}} \int_{\Gamma} (-t^2 + st - s^2 - b)
    e^{\frac{1}{4} t^4 + \frac{b}{2} t^2 -\frac{1}{4} s^4 - \frac{b}{2} s^2} ds dt \\
    & = \lim_{\varepsilon \to 0} \left[
    \frac{1}{2\pi i} \int_{\Sigma_{\varepsilon}} \int_{\Gamma}
        \frac{1}{s-t} e^{\frac{1}{4} t^4 + \frac{b}{2} t^2} \frac{\partial}{\partial s}
        \left[ e^{ -\frac{1}{4} s^4 - \frac{b}{2} s^2}\right] ds dt \right. \\
     & \qquad \qquad \left. + \frac{1}{2\pi i} \int_{\Gamma} \int_{\Sigma_{\varepsilon}}
        \frac{1}{s-t} e^{-\frac{1}{4} s^4 - \frac{b}{2} s^2} \frac{\partial}{\partial t}
        \left[ e^{ \frac{1}{4} t^4 + \frac{b}{2} t^2}\right] dt ds \right]
     \end{aligned}
    \]
If we now do an integration by parts, integrated terms at $\pm \varepsilon$ appear from
the second double integral. The other terms vanish and the result is
\[ \begin{aligned}
   {}[p,q] & =
    \lim_{\varepsilon \to 0}
        \frac{1}{2\pi i} \int_{\Gamma}
            \left[\frac{1}{s+\varepsilon} - \frac{1}{s-\varepsilon} \right]
            e^{-\frac{1}{4} s^4 - \frac{b}{2} s^2}
            e^{\frac{1}{4} \varepsilon^4 + \frac{b}{2} \varepsilon^2} ds
            \end{aligned} \]
Now we deform $\Gamma$ so that instead of the vertical segment $[-i\delta, i\delta]$
it contains the semi-circle $|s| = \delta$, $\Re s > 0$. Then we pick up a
residue contribution from $s = \varepsilon$ which is equal to $1$. The remaining
integral vanishes in the limit $\varepsilon \to 0$, so that we find
$[p,q] =1$, as claimed by the lemma.
\end{proof}

Lemma \ref{GammaSigma} allows us to compute $\widetilde{\Phi}^{-1}$
explicitly. We claim that for $j=1,2,3$,
\begin{equation} \label{defqj}
     q_j(z)  = \frac{1}{2\pi i} \int_{\Sigma_j} e^{\frac{1}{4}t^4 + \frac{b}{2}t^2 +itz} dt,
\end{equation}
where $\Sigma_1$ is a contour in the left half-plane from
$e^{-3\pi i/4}\infty$ to $e^{3\pi i/4} \infty$,
$\Sigma_2$ is a contour in the upper half-plane from
$e^{\pi i/4}\infty$ to $e^{3 \pi/4} \infty$, and
$\Sigma_3$ is a contour in the lower half-plane from
$e^{-3\pi i/4}\infty$ to $e^{-\pi i/4}\infty$.
Indeed, with these contours $\Sigma_j$, and taking note of the
definition and orientation of $\Gamma_0$, $\Gamma_1$, and $\Gamma_4$
in (\ref{pi2}), we easily get from Lemma \ref{GammaSigma} that
the relations (\ref{pqrelations}) hold.
Thus for $q_0 = q_2 + q_3$ we find that $q_0 = - i q$ where
$q$ is defined as in (\ref{Pearceyintegrals}).
Since $p_0 = 2 \pi p$, it is then easy to check that
the formula (\ref{Kcusp6}) for the kernel
is equivalent to the formula (\ref{Pearceykernel}) in the
statement of the theorem. This completes the proof of
Theorem \ref{maintheo}.

\appendix

\section{Proof of Proposition \ref{Rz}}\label{App_A}

Let $\Sigma_R$ be the contour depicted on Figure \ref{fig7}, with
orientation from the left to the right and in the positive
direction on the circles. As usual, we will assume that the minus
side of the contour is on the right.

By a simple arc on $\Sigma_R$ we will mean a connected,

relatively open, with respect to $\Sigma_R$, subset
$\Sigma_R^0\subset\Sigma_R$, which
does not contain any triple point of $\Sigma_R$, a point where three
curves meet. By $L^2(\Sigma_R)$ we will mean, as usual, the space of measurable
functions with
\begin{equation}\label{Rz0}
\| f\|_2=\left(\int_{\Sigma_R} |f|^2|dz|\right)^{\frac{1}{2}}<\infty.
\end{equation}

We have the following general proposition.

\begin{prop}  \label{Rz1}
Suppose that a $3\times 3$ matrix-valued
function $v(z)$, $z\in\Sigma_R$, belongs to $L^2(\Sigma_R)$ and
it is Lipschitz on some simple arc $\Sigma_R^0\subset\Sigma_R$.
Suppose also that on $\Sigma_R^0$, $v(z)$ solves the
equation
\begin{equation}  \label{Rz2}
v(z)=I-\frac{1}{2\pi i}\int_{\Sigma_R}
\frac {v(s)j_R^0(s)}{z_- -s}\,ds,\quad z\in\Sigma_R^0,
\end{equation}
where $z_-$ means the value of the limit of the integral from the minus side,
and $j_R=I+j_R^0$. Then
\begin{equation}\label{Rz3}
R(z)=I-\frac{1}{2\pi i}\int_{\Sigma_R}
\frac {v(s)j_R^0(s)}{z-s}\,ds,\quad z\in\C\setminus\Sigma_R,
\end{equation}
satisfies on $\Sigma_R^0$ the jump condition,
\begin{equation}\label{Rz3a}
R_+(z)=R_-(z)j_R(z),\qquad z\in\Sigma_R^0.
\end{equation}
\end{prop}

\begin{proof} From (\ref{Rz2}), (\ref{Rz3}),
\begin{equation}\label{Rz4}
R_-(z)=v(z),\quad z\in\Sigma_R.
\end{equation}
By the jump property of the Cauchy transform,
\begin{equation}\label{Rz5}
R_+(z)-R_-(z)=v(z)j_R^0(z)=R_-(z)j_R^0(z),
\end{equation}
hence $R_+(z)=R_-(z)j_R(z)$.
Proposition \ref{Rz1} is proved.
\end{proof}

We will solve equation (\ref{Rz2}) by the series,
\begin{equation}\label{Rz6}
v(z)=v_0(z)+v_1(z)+v_2(z)+\dots,
\end{equation}
where
\begin{equation}\label{Rz7}
v_0(z)=I;\qquad v_j(z)=-\frac{1}{2\pi i}\int_{\Sigma_R}
\frac {v_{j-1}(s)j_R^0(s)}{z_- -s}\,ds,\quad z\in\Sigma_R,
\qquad j\ge 1.
\end{equation}
We will inductively estimate $v_j(z)$. We begin with
some general definitions and results.

Introduce the operators
\begin{equation}\label{Rz8}
C^{\pm}_\Ga v(z)=-\frac{1}{2\pi i}\int_{\Ga}
\frac {v(s)}{z_\pm -s}\,ds,\quad z\in\Ga,
\end{equation}
where $\Ga$ is a contour on the complex plane. We assume that
$v$ is Lipschitz and $L^2$ integrable if $\Ga$ is unbounded. We have that
\begin{equation}\label{Rz9}
C^+_\Ga-C^-_\Ga={\rm Id}
\end{equation}
and
\begin{equation}\label{Rz10}
C^+_\Ga+C^-_\Ga=C_\Ga=-\frac{1}{\pi i}{\rm v.p.}\int_{\Ga}
\frac {v(s)}{z -s}\,ds,\quad z\in\Ga.
\end{equation}
Suppose that the contour $\Ga$ is given by the parametric equations,
\begin{equation}\label{Rz11}
\Ga=\{ x=t,\; y=\f(t),\;-\infty< t<\infty\},
\end{equation}
where $\f$ is uniformly Lipschitz,
so that there exists $M\ge 0$ such that
\begin{equation}\label{Rz12}
|\f(x)-\f(y)|\le M|x-y|.
\end{equation}
Then as shown in \cite{CMM}, there exists an absolute constant $K_0$ such that
\begin{equation}\label{Rz13}
\| C_\Ga f\|_2\le K_0(1+M)^{10}\| f\|_2,
\end{equation}
where
\begin{equation}\label{Rz14}
\| f\|_2=\left(\int_\Ga |f|^2|dz|\right)^{\frac{1}{2}}.
\end{equation}
This implies similar estimates for $C^\pm_\Ga$. If $\Ga_0\subset\Ga$
then
\begin{equation}\label{Rz15}
C_{\Ga_0}=PC_{\Ga}P,\qquad Pf=\chi_{\Ga_0}f,
\end{equation}
hence
\begin{equation}\label{Rz16}
\| C_{\Ga_0}\|_2\le \| C_{\Ga}\|_2.
\end{equation}
Therefore, estimate (\ref{Rz13}) holds, with the same constant, for any contour
\begin{equation}\label{Rz17}
\Ga=\{ x=t,\; y=\f(t),\; a< t<b \}.
\end{equation}
Furthermore, it holds, with the same constant, for any complex linear transformation of
contour (\ref{Rz17}). Let us denote
by $\mathcal G_M$ the set of all contours which can be obtained by a complex linear
transformation from a contour (\ref{Rz17}), where $\f$ satisfies (\ref{Rz12})
and is differentiable.
Observe that any interval of a straight line
 belongs to $\mathcal G_0$, and any circular arc of
angular measure less or equal $\frac{\pi}{2}$ belongs  to $\mathcal G_1$.

Suppose now that $\Ga=\Ga_1\cup\ldots\cup\Ga_m$ is a piecewise contour such that
\begin{enumerate}
\item
$\Ga_j$ belongs to $\mathcal G_M$, $j=1,\ldots,m$;
\item
the closed contours,
 $\overline{\Ga_j}$ and $\overline{\Ga_k}$, $j\not= k$, can intersect
only at their end-points;
\item
if $\overline{\Ga_j}$ and $\overline{\Ga_k}$ intersect  then
the angle between them at the intersection point is positive,
\begin{equation}\label{Rz18a}
\angle(\Ga_j,\Ga_k)>\ep>0.
\end{equation}
\item
if $\Ga_j$ and $\Ga_k$, $j\not= k$,
 are two infinite contour then they "well diverge" at infinity,
so that there exists a constant $c>0$ such that
\begin{equation}\label{Rz18}
|\ga_j(s)-\ga_k(t)|\ge c(|s|+|t|),
\end{equation}
where $\ga_j,\;\ga_k$ are the parametric equations of the contours
$\Ga_j,\;\Ga_k$, induced by parametrization (\ref{Rz11}).
\end{enumerate}

\begin{theo} \label{L2}
If $\Ga$ is a piecewise contour which satisfies conditions (1)--(4), then $ C_\Ga$
is bounded in $L^2$, and
 $\| C_\Ga\|_2$ is estimated from above by  a constant which depends only
on the Lipschitz constants $M_j$ of the contours $\Ga_j$, $j=1,\ldots,m$,
and on the constants $\ep$ and $c$ of conditions (\ref{Rz18a}), (\ref{Rz18}).
\end{theo}

\begin{proof} We have to prove that for some $K_1>0$,
\begin{equation}\label{Rz19}
|(C_\Ga f,g)|\le K_1\|f\|_2\| g\|_2.
\end{equation}
To that end, it is sufficient to prove that for some $K_2>0$,
\begin{equation}\label{Rz20}
|(C_\Ga (\chi_{\Ga_j}f),\chi_{\Ga_k} g)|\le K_2\|f\|_2\| g\|_2,
\qquad 1\le j,k\le m.
\end{equation}
For $j=k$, it follows from estimate (\ref{Rz12}) applied to a linear
transformation of $\Ga_j$. For $j\not=k$ it follows from (\ref{Rz18a}),
(\ref{Rz18}), and the estimate,
\begin{equation}\label{Rz21}
\frac{1}{\pi}\int_0^\infty\int_0^\infty \frac{|f(s)g(t)|dsdt}{s+t}
\le \|f\|_2\| g\|_2.
\end{equation}
Theorem  \ref{L2} is proved.
\end{proof}

When applied to the contour $\Sigma_R$, Theorem \ref{L2} gives that there exists
a constant $K$, independent of $n$, such that
\begin{equation}\label{Rz22}
\| C_{\Sigma_R} \|_2\le K.
\end{equation}
By (\ref{Rz9}), (\ref{Rz10}) this implies that
\begin{equation}\label{Rz23}
\| C_{\Sigma_R}^\pm \|_2\le K_0=\frac{K+1}{2}\,.
\end{equation}
From (\ref{Rz7}) we have that
\begin{equation}\label{Rz24}
v_j=C_{\Sigma_R}^{-}(j_R^0v_{j-1}). \qquad j\ge 1.
\end{equation}
Since
\begin{equation}\label{Rz25}
\| j_R^0v_{j-1} \|_2\le \| j_R^0\|_C\,\| v_{j-1}\|_2
\end{equation}
and
\begin{equation}\label{Rz26}
\| j_R^0 \|_C=\sup_{z\in \Sigma_R}| j_R^0(z)|\le K_1 n^{-\frac{1}{6}},
\end{equation}
we obtain the recursive estimate,
\begin{equation}\label{Rz27}
\| v_j \|_2\le K n^{-\frac{1}{6}}\| v_{j-1}\|_2,\qquad K=K_0K_1.
\end{equation}
For $v_1$ we have that
\begin{equation}\label{Rz28}
\| v_1 \|_2=\| C^-_{\Sigma_R}(j_R^0)\|_2\le K_0 \| j_R^0\|_2\le K_3 n^{-\frac{1}{6}-\frac{1}{8}}.
\end{equation}
Thus,
\begin{equation}\label{Rz29}
\| v_j \|_2\le K_3(K n^{-\frac{1}{6}})^j n^{-\frac{1}{8}}.
\end{equation}
This implies the convergence of series (\ref{Rz6}) in $L^2$,
for large $n$.
Let us discuss analytic properties of the functions $v_j$.

Denote
\begin{equation}\label{Rz30}
\Sigma_R=\bigcup_{l=1}^{16} \Sigma_R^l,
\end{equation}
the partition of the contour $\Sigma_R$ (see Figure \ref{fig7}) into 16 simple
arcs. Fix any $\ep>0$. Let $z_0$ be any point on $\Sigma_R^l$ such that
the distance from $z_0$ to the end-points of $\Sigma_R^l$ is bigger
than
\begin{equation}\label{Rz30a}
\ep_n=\ep n^{-\frac{1}{4}}.
\end{equation}
The function
$j_R^0(z)$ can be analytically continued from $\Sigma_R^l$ to
the $\ep_n$-neighborhood of the point $z_0$,
\begin{equation}\label{Rz30b}
D(z_0,\ep_n)=\{z:\; \dist(z,z_0)<\ep_n\}.
\end{equation}
This implies that
\begin{equation}\label{Rz31}
v_1(z)=-\frac{1}{2\pi i}\int_{\Sigma_R}
\frac {j_R^0(s)}{z_- -s}\,ds
\end{equation}
can be also analytically continued from $\Sigma_R^l$ to $D(z_0,\ep_n)$,
because we can deform the contour of integration, $\Sigma_R$.
Then, inductively, we can analytically continue $v_j(z)$ from $\Sigma_R^l$ to
$D(z_0,\ep_n)$, by deforming the contour of integration in
(\ref{Rz7}). Observe that on the deformed contour we have
the $L^2$-estimate, (\ref{Rz29}), hence by the Cauchy-Schwarz
inequality we obtain that
\begin{equation}\label{Rz32}
|v_j(z)|\le K_4 \ep^{-1}n^{\frac{1}{4}}(K n^{-\frac{1}{6}})^{(j-1)} n^{-\frac{1}{8}},
\qquad z\in D(z_0,\ep_n/2).
\end{equation}
This proves the convergence of series (\ref{Rz6}) in the
neighborhood $D(z_0,\ep_n/2)$ to an analytic $v(z)$.
Thus, $v(z)$ is analytic on $\Sigma_R$ outside of the triple points.

Observe that the function $R(z)$ defined by formula (\ref {Rz3}), denote it for a moment
$\tilde R(z)$, coincides
with $R(z)$ defined by (\ref{defR}). Indeed, both $\tilde R(z)$ and $R(z)$ solve
the same RH problem, and if $z_0$ is any triple point of $\Sigma_R$,
 then in some neighborhood of $z_0$,
\begin{equation}\label{Rz32a}
|\tilde R(z)|\le C|z-z_0|^{-\frac{1}{2}},\qquad |R(z)|\le C,
\end{equation}
for some $C>0$. For $\tilde R(z)$ it follows from (\ref{Rz3}) by the Cauchy-Schwarz inequality,
and for $R(z)$ it is obvious from (\ref{defR}). If we consider now
\begin{equation}\label{Rz32b}
X(z)=\tilde R(z) R(z)^{-1},
\end{equation}
then $X(z)$ has no jumps on $\Sigma_R$ and in a neighborhood
of the triple points it satisfies the estimate
\begin{equation}\label{Rz32c}
|X(z)|\le C|z-z_0|^{-\frac{1}{2}}.
\end{equation}
Therefore, the triple points are removable singularities and $X(z)$ is analytic on $\C$.
Also, $X(\infty)=I$, hence $X(z)=I$ everywhere on $\C$, and $\tilde R(z)=R(z)$.
Now we can estimate $R(z)$.

From (\ref{Rz3}),
\begin{equation}\label{Rz33}
R(z)=I+\sum_{j=0}^\infty R_j(z),
\qquad
R_j(z)=-\frac{1}{2\pi i}\int_{\Sigma_R}
\frac {v_j(s)j_R^0(s)}{z-s}\,ds.
\end{equation}
Suppose that
\begin{equation}\label{Rz34}
 \dist(z,\Sigma_R)>0.1n^{-\frac{1}{4}}.
\end{equation}
Then,
\begin{equation}\label{Rz35}
|R_0(z)|=\frac{1}{2\pi }\left|\int_{\Sigma_R}
\frac {j_R^0(s)}{z-s}\,ds\right|\le \frac{K_0 n^{-\frac{1}{6}}}{1+|z|}\,,
\end{equation}
and by (\ref{Rz29}),
\begin{equation}\label{Rz36}
|R_j(z)|\le \frac{K_1n^{\frac{1}{4}}(Kn^{-\frac{1}{6}})^{j+1}n^{-\frac{1}{8}}}{1+|z|}
=\frac{K_1(Kn^{-\frac{1}{6}})^{j}n^{-\frac{1}{24}}}{1+|z|}\,,\qquad j\ge 1.
\end{equation}
By summing all these inequalities from $j=0$ to $\infty$, we obtain that
\begin{equation}\label{Rz37}
R(z)=I+O\left(\frac{n^{-\frac{1}{6}}}{1+|z|}\right),\qquad \dist(z,\Sigma_R)>
0.1n^{-\frac{1}{4}}.
\end{equation}
In fact, the restriction $\dist(z,\Sigma_R)>
0.1n^{-\frac{1}{4}}$ is not essential, because we can deform the contour
$\Sigma_R$. This completes the proof of Proposition \ref{Rz}.


\begin{thebibliography}{99}

\bibitem{AvM}
    M. Adler and P. van Moerbeke,
    PDE's for the Gaussian ensemble with external source
    and the Pearcey distribution,
    arxiv:math.PR/0509047.

\bibitem{ABK} A.I. Aptekarev, P.M. Bleher, and A.B.J. Kuijlaars,
    Large $n$ limit of Gaussian random matrices with external source, part II,
    Commun. Math. Phys. 259 (2005), 367--389.

\bibitem{ABVA}
    A.I. Aptekarev, A. Branquinho, and W. Van Assche,
    Multiple orthogonal polynomials for classical weights,
    Trans. Amer. Math. Soc. 355 (2003), 3887--3914.

\bibitem{BHo}
    M.V. Berry and C.J. Howls,
    Hyperasymptotics for integrals with saddles,
    Proc. Roy. Soc. London Ser. A 434 (1991), 657--675.

\bibitem{BI} P. Bleher and A. Its,
    Double scaling limit in the random matrix model. The Riemann-Hilbert approach,
    Commun. Pure Appl. Math. 56 (2003), 433--516.

\bibitem{BK1} P.M. Bleher and A.B.J. Kuijlaars,
    Random matrices with external source and multiple orthogonal polynomials,
    Int. Math. Research Notices 2004, no 3 (2004), 109--129.

\bibitem{BK2} P.M. Bleher and A.B.J. Kuijlaars,
    Large $n$ limit of Gaussian random matrices with external source, part I,
    Commun. Math. Phys. 252 (2004), 43--76.

\bibitem{BK3} P.M. Bleher and A.B.J. Kuijlaars,
    Integral representations for multiple Hermite and multiple Laguerre polynomials,
    Ann. Inst. Fourier  55 (2005), 2001--2004.

\bibitem{BH1} E. Br\'ezin and S. Hikami,
    Universal singularity at the closure of a gap in a random matrix theory,
    Phys. Rev. E 57 (1998), 4140--4149.

\bibitem{BH2} E. Br\'ezin and S. Hikami,
    Level spacing of random matrices in an external source,
    Phys. Rev. E 58 (1998), 7176--7185.

\bibitem{CK} T. Claeys and A.B.J. Kuijlaars,
    Universality of the double scaling limit in random matrix models,
    arxiv:math-ph/0501074,
    to appear in Commun. Pure Appl. Math.

\bibitem{CKV} T. Claeys, A.B.J. Kuijlaars, and M. Vanlessen,
     Multi-critical unitary random matrix ensembles and the
     general Painlev\'e II equation,
     arxiv:math-ph/0508062.

\bibitem{CMM} R.R. Coifman, A. McIntosh, and Y. Meyer,
     L'int\'egrale de Cauchy d\'efinit un op\'erateur born\'e sur $L^2$
     pour les courbes Lipschitziennes,
     Ann. Math.  116 (1982), 361--387.

\bibitem{DK} E. Daems and A.B.J. Kuijlaars,
    A Christoffel-Darboux formula for multiple orthogonal polynomials,
    J. Approx. Theory 130 (2004), 190--202.

\bibitem{Dei} P. Deift,
    Orthogonal Polynomials and Random Matrices: A Riemann-Hilbert approach,
    Courant Lecture Notes in Mathematics Vol. 3, Amer. Math. Soc.,
    Providence R.I. 1999.

\bibitem{DKMVZ1} P. Deift, T. Kriecherbauer,
    K.T-R McLaughlin, S. Venakides, and X. Zhou,
    Uniform asymptotics of polynomials orthogonal with respect to varying
    exponential weights and applications to universality questions in
    random matrix theory, Commun. Pure Appl. Math. 52 (1999), 1335--1425.

\bibitem{DKMVZ2} P. Deift, T. Kriecherbauer,
    K.T-R McLaughlin, S. Venakides, and X. Zhou,
    Strong asymptotics of orthogonal polynomials with respect to
    exponential weights, Commun. Pure Appl. Math 52 (1999), 1491--1552.

\bibitem{DZ} P. Deift and X. Zhou,
    A steepest descent method for oscillatory Riemann-Hilbert problems.
    Asymptotics for the MKdV equation,
    Ann. Math. 137 (1993), 295--368.

\bibitem{Miy} T. Miyamoto,
    On an Airy function of two variables,
    Nonlinear Anal. 54 (2003), 755--772.

\bibitem{OR} A. Okounkov and N. Reshetikhin,
    Random skew plane partitions and the Pearcey process,
    arxiv:math.CO/0503508.

\bibitem{PK} R.B. Paris and D. Kaminski,
    Asymptotics and Mellin-Barnes integrals,
    Cambridge University Press, Cambridge, 2001.

\bibitem{Pas} L. Pastur,
    The spectrum of random matrices (Russian),
    Teoret. Mat. Fiz. 10 (1972), 102--112.

\bibitem{Pearcey} T. Pearcey,
    The structure of an electromagnetic field in the neighborhood of a cusp of a caustic,
    Philos. Mag. 37 (1946), 311--317.

\bibitem{ST} E.B. Saff and V. Totik,
    Logarithmic Potentials with External Field,
    Springer-Verlag, 1997.

\bibitem{TW} C. Tracy and H. Widom,
    The Pearcey process,
    Commun. Math. Phys. 263 (2006), 381--400.

\bibitem{VAC} W. Van Assche and E. Coussement,
    Some classical multiple orthogonal polynomials,
    J. Comput. Appl. Math. 127 (2001), 317--347.

\bibitem{VAGK} W. Van Assche, J. Geronimo, and A.B.J. Kuijlaars,
    Riemann-Hilbert problems for multiple orthogonal polynomials,
    In: Special Functions 2000 (J. Bustoz et al., eds.), Dordrecht,
    Kluwer, 2001, pp. 23--59.

\bibitem{ZJ1} P. Zinn-Justin,
    Random Hermitian matrices in an external field,
    Nucl. Phys. B 497 (1997), 725--732.

\end{thebibliography}
\end{document}